\documentclass{JHEP3} 

\usepackage{graphicx}
\usepackage{amsmath, amsfonts}
\usepackage{epsfig}

\usepackage[normalem]%
{ulem}  

\newcommand{\non}{\nonumber\\}

\newcommand{\be}{\begin{equation}}
\newcommand{\ee}{\end{equation}}
\newcommand{\bea}{\begin{eqnarray}}
\newcommand{\eea}{\end{eqnarray}}
\newcommand{\ba}[1]{\begin{array}{#1}}
\newcommand{\ea}{\end{array}}

\newcommand{\bm}[1]{\mbox{\boldmath${#1}$}}

\title{Anomalies and the chiral magnetic effect in the Sakai-Sugimoto model}

\author{Anton Rebhan, Andreas Schmitt, Stefan A. Stricker\\
Institut f\"{u}r Theoretische Physik, Technische Universit\"{a}t Wien, 1040 Vienna, Austria\\ 
\\
 \email{rebhana@hep.itp.tuwien.ac.at}\\
 \email{aschmitt@hep.itp.tuwien.ac.at}\\
  \email{stricker@hep.itp.tuwien.ac.at}}

\abstract{In the chiral magnetic effect an imbalance in the number of left- and right-handed quarks gives rise to an electromagnetic 
current parallel to the magnetic field produced in noncentral heavy-ion collisions. The chiral imbalance may be 
induced by topologically nontrivial gluon configurations 
via the QCD axial anomaly, while the resulting electromagnetic current itself is a consequence of the QED anomaly. 
In the Sakai-Sugimoto model, which in a certain limit 
is dual to large-$N_c$ QCD, we discuss the proper implementation of the QED axial anomaly,
the (ambiguous) definition of chiral currents, and the calculation
of the chiral magnetic effect. 
We show that this model correctly contains the so-called consistent anomaly,
but requires the introduction of a (holographic) finite counterterm to yield the
correct covariant anomaly. Introducing net chirality through an axial chemical potential, we find a nonvanishing 
vector current only before including this counterterm. This seems to imply the absence of the chiral magnetic effect in 
this model. On the other hand, for a conventional quark chemical potential and large
magnetic field, which is of interest in the physics of compact stars,
we obtain a nontrivial result for the axial current that is
in agreement with previous calculations and known exact results for QCD.
}


\begin{document}

\section{Introduction}
\label{intro}

Topologically charged gauge field configurations in QCD generate chirality due to the nonabelian axial anomaly. 
In the presence of a magnetic field, this chirality, i.e., an imbalance in the number of right- and left-handed quarks,
has been predicted to generate an electromagnetic current parallel to the applied magnetic field. 
This is a consequence of the QED axial anomaly and has been termed chiral magnetic effect \cite{Kharzeev:2007jp,Fukushima:2008xe,Kharzeev:2009pj}. 
As a result, electric charge separation may occur 
in noncentral heavy-ion collisions, 
where magnetic fields up  to $10^{17}\,{\rm G}$ can be generated temporarily,
and corresponding experimental evidence has in fact been reported 
in refs.\ \cite{Voloshin:2008jx,Abelev:2009uh} (see however ref.\ \cite{Wang:2009kd}). 

In a simplified picture, one may study the induced current for a static magnetic field. The generalization to time-dependent magnetic fields,
as produced in heavy-ion collisions, in principle amounts to computing a frequency-dependent 
conductivity \cite{Kharzeev:2009pj,Yee:2009vw}. However, the observed 
charge separation is proportional to the zero-frequency limit \cite{Kharzeev:2009pj}. In this paper, the currents we compute always 
correspond to the zero-frequency limits of the conductivities. 

Another simplification of the highly nontrivial scenario of a heavy-ion 
collision is to mimic the 
(event-by-event) topologically induced chirality by a nonzero axial chemical potential $\mu_5$, 
the difference of right- and left-handed chemical potentials. The resulting current is a vector current proportional 
to $\mu_5$. In a more general setup, although negligible in the heavy-ion context, 
one may also include a quark chemical potential $\mu$, which is the same for right- and left-handed fermions. Again via a 
nonzero magnetic field, an axial current is generated in this case \cite{Son:2004tq,Metlitski:2005pr}. This effect may be of relevance for the physics of compact stars \cite{Gorbar:2009bm}, 
where strongly interacting matter can reach densities of 
several times nuclear ground state density, and (surface) magnetic fields up to $10^{15}\, {\rm G}$ have been measured, indicating the 
possibility of even higher magnetic fields in the interior. 
Also the direct high-density
analogue of the chiral magnetic effect has been studied in the context of
neutron star physics \cite{Charbonneau:2009ax}.

In the present paper, we apply a strong-coupling approach, based on the AdS/CFT correspondence \cite{Maldacena:1997re,Gubser:1998bc,Witten:1998qj}, 
to compute both kinds of currents. We use a general setup to account for nonzero temperatures, relevant
in the context of heavy-ion collisions, as well as for nonzero quark chemical potentials, 
relevant in the astrophysical context. Besides the chirally symmetric phase we also consider the chirally broken phase
which is important in both contexts: heavy-ion collisions are expected to probe the region of the QCD chiral phase transition; in quark matter at densities present in 
compact stars, chiral symmetry may also be spontaneously broken, for example in the color-flavor locked phase \cite{Alford:1998mk}.

We use the Sakai-Sugimoto model \cite{Sakai:2004cn,Sakai:2005yt}, which is sometimes called ``holographic QCD'' since 
in the limit of small 't Hooft coupling it provides a string theory dual to \mbox{large-$N_c$ QCD}. However, we work in the opposite, 
strongly coupled, limit where the simple gravity approximation can be employed but where the model is no longer dual to QCD. 
The model still yields interesting qualitative predictions especially in view of the strong-coupling nature of both 
contexts mentioned above, i.e., in QCD at large (but not asymptotically large) temperature and small quark chemical potential and QCD at large 
(but not asymptotically large) quark chemical potential and small temperature. 

The Sakai-Sugimoto model is particularly suited for our purpose since it has a well-defined concept for chirality and the chiral 
phase transition. It is straightforward to introduce right- and left-handed chemical potentials independently. Several previous works 
have considered currents in a magnetic field at nonzero chemical potentials in this model 
\cite{Bergman:2008qv,Thompson:2008qw,Rebhan:2008ur,Lifschytz:2009si}. 
The purpose of the present paper is two-fold. The physical motivation is to extend these calculations to the currents  
relevant for the chiral magnetic effect, and to compare our strong-coupling results to the  
weak-coupling results \cite{Fukushima:2008xe} as well as the lattice results \cite{Buividovich:2009wi} in the existing literature. 
There is also a more theoretical purpose of our work, addressing certain fundamental properties of the Sakai-Sugimoto model. 
We discuss in detail how to implement the covariant QED anomaly into the model in order to obtain physically acceptable predictions. 
Moreover, we elaborate on an ambiguity in the definition of the chiral currents in the presence of a Chern-Simons term that has been observed previously \cite{Bergman:2008qv,Rebhan:2008ur,Lifschytz:2009si} (see also \cite{Hata:2008xc}).

Our paper is organized as follows. We start with a brief introduction into the model and a general discussion of the currents, in particular the appearance of consistent and covariant anomalies, in sec.\ \ref{sec:anomalies}. In sec.\ \ref{sec:background} we discuss the solution of the equations of
motion in the presence of background magnetic and electric fields. We present analytical solutions
for the chirally broken phase, sec.\ \ref{sec:solution1}, and the chirally symmetric phase, sec.\ \ref{sec:solution2}. We then discuss the 
ambiguity of the currents, defined on the one hand via the general definition from sec.\ \ref{sec:anomalies}, and on the other hand 
from the thermodynamic potentials obtained in sec.\ \ref{sec:background}. In sec.\ \ref{sec:cme} we present our results for 
the axial and vector currents and give our conclusions in sec.\ \ref{sec:conclusions}.

\section{Anomalies in the Sakai-Sugimoto model}
\label{sec:anomalies}

\subsection{Brief introduction into the model}
\label{sec:brief}

The Sakai-Sugimoto model is based on ten-dimensional type-IIA string theory, with a background geometry given by 
$N_c$ D4-branes. They span four-dimensional space-time $(\tau, {\bf x})$ and a fifth extra dimension $x_4$ compactified on a circle
whose circumference is parametrized by the Kaluza-Klein mass $M_{\rm KK}$, $x_4\equiv x_4 + 2\pi/M_{\rm KK}$. Through this compactified dimension
and antisymmetric boundary conditions for fermions
supersymmetry is completely broken. Left- and right-handed chiral fermions are 
introduced by adding $N_f$ D8- and $N_f$ $\overline{\rm D8}$-branes which extend in all dimensions except $x_4$. In this
compact direction, they are separated by a distance $L\in[0,\pi/M_{\rm KK}]$. 
For more details about the setup of the model and the explicit form of the background
metric, including the holographic direction $z$ and a four-sphere $S^4$, 
see the original papers by Sakai and Sugimoto, refs.\ \cite{Sakai:2004cn,Sakai:2005yt}. 
We employ the probe brane approximation, i.e., the background geometry is assumed to be unaltered by the flavor branes.  
This is a good approximation for $N_f\ll N_c$.

There are two possible background geometries. One, interpreted as the confined phase, has a cigar-shaped $(x_4,z)$ subspace, ending 
at the tip $z=0$, and a cylinder-shaped $(\tau,z)$ subspace, where $\tau$ is Euclidean time on a circle with circumference given 
by the inverse temperature, $\tau\equiv \tau+1/T$. In the other geometry, interpreted as the deconfined phase, $x_4$ and $\tau$ 
exchange their roles, such 
that the $(x_4,z)$ subspace is cylinder-shaped while the $(\tau,z)$ subspace is cigar-shaped, corresponding to a geometry with black-hole horizon after analytical continuation $i\tau\to
t$.

Chiral symmetry breaking is realized in the model as follows. A $U(N_f)$ gauge symmetry on the flavor branes corresponds to a global 
$U(N_f)$ at the boundary. Therefore, the bulk gauge symmetries on the D8- and $\overline{\rm D8}$-branes can be interpreted as 
left- and right-handed flavor symmetry groups in the dual field theory. 
The Chern-Simons term accounts for the axial anomaly of QCD, such that one is left with 
the chiral group $SU(N_f)_L\times SU(N_f)_R$ and the vector part $U(1)_V$.  There is no explicit breaking of this group since the 
model only contains massless quarks. Spontaneous chiral symmetry breaking is realized when the D8- and $\overline{\rm D8}$-branes connect
in the bulk. They always connect in the confined phase, where the $(x_4,z)$ subspace
is singly connected. 
Whether they connect in the deconfined phase depends on the separation $L$ of the 
D8- and $\overline{\rm D8}$-branes in the extra dimension $x_4$. Here we shall always consider maximally separated branes, $L=\pi/M_{\rm KK}$.
With this choice the flavor branes necessarily extend to the black-hole horizon and
thus never connect in the deconfined phase. Consequently, the deconfinement and chiral phase transitions are identical and happen at a critical 
temperature $T_c=M_{\rm KK}/(2\pi)$, i.e., when the radii of the $\tau$ and $x_4$ circles are equal. We shall always use the terminology of chirally symmetric and chirally broken phases. 
This is equivalent to speaking about deconfined and confined phases, but more appropriate in our context because we are interested 
in the interplay of a magnetic field with chirality. In the probe brane approximation, the chiral/deconfinement phase transition is given 
solely by the background geometry and is not affected by the gauge fields on the flavor branes. In particular, it does not depend on the 
chemical potential, which, at least for the deconfinement phase transition, is in accordance with expectations for large-$N_c$ QCD \cite{McLerran:2007qj}. 

\subsection{Action, equations of motion, and currents}
\label{sec:action}

In this section we discuss the general equations of the model in the broken phase where the D8- and $\overline{\rm D8}$-branes are connected. 
The equations for the symmetric phase are very similar and shall be given later where necessary. The D-brane action consists of a Dirac-Born-Infeld
(DBI) and a Chern-Simons (CS) part. We approximate the DBI action by the Yang-Mills (YM) action which is a good approximation
for small magnetic fields. The use of the YM action greatly simplifies the treatment since the equations
of motion then have solutions which can be given in an almost entirely analytical way. Throughout the paper we shall work with one quark flavor, 
$N_f=1$. The currents we compute are expected to be simple sums over quark flavors, 
each flavor contributing in the same way, distinguished only by its electric charge. This is rather obvious in the chirally symmetric phase. In the
chirally broken phase, the flavor contributions may be more complicated in the case of charged pion condensation. However, since we work 
at vanishing isospin chemical potential, there is only neutral pion condensation and the different flavor contributions decouple. 

For one quark flavor and the gauge $A_z=0$ the (Euclidean) action
\be
S = S_{\rm YM} + S_{\rm CS} 
\ee
is given by  \cite{Sakai:2005yt}
\begin{subequations} \label{SYMCS}
\bea
S_{\rm YM} &=& \kappa M_{\rm KK}^2 \int d^4 x\int_{-\infty}^{\infty} dz \, 
\left[k(z)F_{z\mu}F^{z\mu}+\frac{h(z)}{2M_{\rm KK}^2}F_{\mu\nu}F^{\mu\nu}\right] \, , \label{SYM}\\
S_{\rm CS} &=& \frac{N_c}{24\pi^2} \int d^4 x\int_{-\infty}^{\infty} dz\, A_\mu F_{z\nu}F_{\rho\sigma}\epsilon^{\mu\nu\rho\sigma} \, , \label{SCS}
\eea
\end{subequations}
with Greek indices running over $\mu,\nu,\ldots = 0,1,2,3$. 
Our convention for the epsilon tensor is $\epsilon_{0123}=+1$. 
In eq.\ (\ref{SYMCS}) we have defined the metric functions  
\be
k(z)\equiv 1+z^2 \, , \qquad h(z)\equiv (1+z^2)^{-1/3} \, , 
\ee
and the dimensionless constant
\be \label{kappa}
\kappa \equiv \frac{\lambda N_c}{216\pi^3} \, ,
\ee
where $\lambda$ is the 't Hooft coupling. The integration over the four-sphere has already been done, and we are left with the integral 
over space-time $(\tau,{\bf x})$ and the holographic coordinate $z$ which extends from the left-handed boundary ($z=+\infty$) over
the tip of the cigar-shaped $(x^4,z)$ subspace ($z=0$) to the right-handed boundary ($z=-\infty$). The coordinate $z$ is dimensionless
and is obtained from the dimensionful coordinate $z$ of ref.\ \cite{Rebhan:2008ur} upon defining $z'=z/u_{\rm KK}$ $(T<T_c)$ 
and then dropping the prime. Here, $u_{\rm KK}=4R^3M_{\rm KK}^2/9$, with $R$ being the curvature radius of the background metric. 
Since we work at finite temperature,  
we need to work in Euclidean space. However, we use Minkowski notation which is more convenient for the following discussion of the anomaly.
More precisely, we start from the Euclidean action with imaginary time $\tau$ and replace $A_0\to iA_0$, after which 
we may write the result using a Minkowski metric with signature $(-,+,+,+)$. The space-time integral is denoted by $d^4x$ for simplicity but
actually is an integral $d\tau\,d^3x$ over imaginary time $\tau$ and three-dimensional space. For the general form of the action, without any gauge choice and for more flavors, see for instance 
refs.\ \cite{Sakai:2004cn,Rebhan:2008ur,Hata:2007mb}. 
The equations of motion for $N_f=1$ are
\begin{subequations} \label{eom}
\bea
\kappa M_{\rm KK}^2\,
\partial_z[k(z)F^{z\mu}]+\kappa
h(z)\partial_\nu F^{\nu\mu} &=&\frac{N_c}{16\pi^2}F_{z\nu}F_{\rho\sigma}
\epsilon^{\mu\nu\rho\sigma} \, , \label{eq1}\\
\kappa M_{\rm KK}^2\,
\partial_\mu[k(z)F^{z\mu}] &=&\frac{N_c}{64\pi^2}F_{\mu\nu}F_{\rho\sigma}\epsilon^{\mu\nu\rho\sigma} \, , \label{eq2}
\eea
\end{subequations}
where the second equation 
is obtained from varying $A_z$ in the action prior to setting $A_z=0$.

Next we introduce the chiral
currents. The usual way is to define them through the variation of the on-shell action with respect to the boundary values
of the gauge fields (see however
ref.~\cite{Hata:2008xc} for a discussion of possible alternatives). 
We thus replace $A_\mu(x,z)\to A_\mu(x,z) + \delta A_\mu(x,z)$ in the action  and keep the 
terms linear in $\delta A_\mu(x,z)$ to obtain
\begin{subequations} \label{currboundary}
\bea
\delta S_{\rm YM} &=& 2\kappa M_{\rm KK}^2\left\{\int d^4x\,  k(z) F^{z\mu}\delta A_\mu \Big|_{z=-\infty}^{z=\infty} 
+\int d^3x \int_{-\infty}^{\infty} dz\, \frac{h(z)}{M_{\rm KK}^2}F^{\nu\mu}\delta A_\mu\Big|_{x_\nu} \right. \non
&& \left. -\, \int d^4x \int_{-\infty}^{\infty} dz\,\left[ \partial_z[k(z)F^{z\mu}] + 
\frac{h(z)}{M_{\rm KK}^2}\partial_\nu F^{\nu\mu}\right] \delta A_\mu \right\} \, , \\
\delta S_{\rm CS} &=& \frac{N_c}{8\pi^2}\left\{ -\frac{1}{3}\int d^4x\, A_\nu F_{\rho\sigma}\delta A_\mu \Big|_{z=-\infty}^{z=\infty}  
-\frac{2}{3} \int d^3x \int_{-\infty}^{\infty} dz\, A_\sigma F_{z\nu}\delta A_\mu \Big|_{x_\rho} \right. \non
&& \left. +\,\int d^4x \int_{-\infty}^{\infty} dz\, F_{z\nu}F_{\rho\sigma}\delta A_\mu 
\right\} \epsilon^{\mu\nu\rho\sigma} \, . 
\eea 
\end{subequations}
In the total variation $\delta S = \delta S_{\rm YM}+\delta S_{\rm CS}$ the bulk terms vanish upon using the equation of motion 
for $A_\mu$ (\ref{eq1}) and we are left with boundary terms only. 
According to the holographic correspondence, we keep only the boundary terms at 
$|z|=\infty$ and drop any terms from space-time infinities. 
This may seem natural but possibly is problematic in our case 
as we shall discuss later after we have implemented our specific ansatz. 
The boundary terms at the holographic boundary
$z=\pm\infty$ lead to the left- and right-handed currents
\be \label{currents}
{\cal J}^\mu_{L/R} \equiv -\frac{\delta S}{\delta A_\mu(x,z=\pm\infty)}= 
\mp\left(2\kappa M_{\rm KK}^2k(z)F^{z\mu} -\frac{N_c}{24\pi^2}\epsilon^{\mu\nu\rho\sigma}
A_\nu F_{\rho\sigma}\right)_{z=\pm\infty} \, ,
\ee
where the first (second) term is the YM (CS) contribution. This result of the currents is in agreement with 
refs.\ \cite{Rebhan:2008ur,Hata:2008xc,Hashimoto:2008zw}, see also \cite{Kim:2008pw,Kim:2008iy}. The overall
minus sign in the definition originates from our use of the Euclidean
action which is minus the Minkowski action,
and the functional derivative is taken with respect to 
the space-time coordinates $x$ (and not also
with respect to the holographic coordinate $z$ plus a subsequent limit $z\to \pm\infty$). 
The currents (\ref{currents}) can also be obtained from 
\be \label{dLdA}
{\cal J}_{L/R}^\mu = \mp \left. \frac{\partial {\cal L}}{\partial\,\partial_zA_\mu}\right|_{z=\pm\infty} \, , 
\ee
in accordance with the usual rules of the gauge/gravity correspondence. 

As already pointed out in ref.\ \cite{Rebhan:2008ur}, 
it is only the YM part of the current, i.e., the first term in eq.\ (\ref{currents}), which appears in the asymptotic
expansion of the gauge fields. From the definition (\ref{currents}) and with $k(z)=1+z^2$ we find
\bea \label{expandYM}
A_\mu(x,z) &=& A_\mu(x,z=\pm\infty)\pm \frac{{\cal J}_{\mu,{\rm YM}}^{L/R}}{2\kappa M_{\rm KK}^2}\frac{1}{z} 
+ {\cal O}\left(\frac{1}{z^2}\right) \, .
\eea
One can also confirm this relation from our explicit results in the subsequent sections. 

\subsection{Consistent and covariant anomalies}
\label{sec:concov}

The divergence of the currents (\ref{currents}) can be easily computed with the help of the equation of motion for $A_z$ (\ref{eq2}). 
One obtains
\be \label{consanomaly}
\partial_\mu{\cal J}^\mu_{L/R} =
\partial_\mu(\mathcal J_{\rm YM}+\mathcal J_{\rm CS})^\mu_{L/R} =
\mp \frac{N_c}{16\pi^2}\left(1-\frac23\right)
 F_{\mu\nu}^{L/R} \widetilde{F}^{\mu\nu}_{L/R} \, ,
\ee
with the left- and right-handed field strengths $F_{\mu\nu}^{L/R}(x) \equiv F_{\mu\nu}(x,z=\pm\infty)$, 
and the left- and right-handed dual field strength tensors $\widetilde{F}^{\mu\nu}_{L/R}=\frac12\,F_{\rho\sigma}^{L/R}\epsilon^{\mu\nu\rho\sigma}$. 
(For notational convenience we use the labels $L$, $R$ and related labels such as $V,A$ sometimes as superscript, sometimes as subscript.) 
With the vector and axial currents
\be \label{JVA}
{\cal J}^\mu \equiv {\cal J}^\mu_R+{\cal J}^\mu_L \, , \qquad {\cal J}^\mu_5 \equiv {\cal J}^\mu_R-{\cal J}^\mu_L \, , 
\ee
and the vector and axial field strengths introduced as $F_{\mu\nu}^R=F_{\mu\nu}^V+F_{\mu\nu}^A$, $F_{\mu\nu}^L=F_{\mu\nu}^V-F_{\mu\nu}^A$, 
eq.\ (\ref{consanomaly}) yields the vector and axial anomalies
\begin{subequations}
\bea
\partial_\mu{\cal J}^\mu &=& \frac{N_c}{12\pi^2} F_{\mu\nu}^V\widetilde{F}^{\mu\nu}_A \, , \\
\partial_\mu{\cal J}^\mu_5 &=& \frac{N_c}{24\pi^2} \left(F_{\mu\nu}^V\widetilde{F}^{\mu\nu}_V+F_{\mu\nu}^A\widetilde{F}^{\mu\nu}_A\right) \, .
\label{consanomaly1b}
\eea
\end{subequations}
The coefficients on the right-hand side 
(which as we saw receive contributions from both
the YM and CS parts of the currents) are in accordance with the standard
field theoretic results for $N_c$ chiral fermionic
degrees of freedom coupled to left and right chiral gauge fields 
\cite{Bardeen:1969md}. 
The above form of the anomaly, which is symmetric in vector and axial-vector gauge fields, is called {\it consistent} anomaly.
If left- and 
right-handed Weyl spinors are treated separately, 
this form of the anomaly arises unambiguously. This is
explained for instance in 
ref.\ \cite{Hill:2006ei}, where left- and right-handed fields are 
separated by an extra dimension. 
This is not unlike our present model and it is thus not surprising that the consistent anomaly arises naturally from the above definition of the
currents. In QED, however, we must require that 
the vector current be strictly conserved, 
even in the presence of axial field strengths.
As was first discussed by Bardeen \cite{Bardeen:1969md}, this
can be achieved by the introduction of a counterterm that
mixes left- and right-handed gauge fields.
Having even parity, Bardeen's counterterm is uniquely given by \cite{Hill:2006ei}   
\be\label{DeltaS}
\Delta S = c\int d^4x (A_\mu^L A_\nu^R F_{\rho\sigma}^L+A_\mu^L A_\nu^R F_{\rho\sigma}^R)\epsilon^{\mu\nu\rho\sigma} \, ,
\ee
where $c$ is a constant determined by requiring a strictly conserved vector current. Because this expression can be naturally written as a (metric-independent) 
integral over a
hypersurface at $|z|=\Lambda\to\infty$ with left- and right-handed fields
concentrated at the respective brane locations, $\Delta S$ can
actually be interpreted as a (finite) counterterm in holographic renormalization.
In particular, it does not change the equations of motion.

To obtain the contribution of Bardeen's counterterm to the chiral currents we 
replace $A_\mu^{L/R}\to A_\mu^{L/R} + \delta A_\mu^{L/R}$ to obtain 
\bea
\delta\Delta S &=& \pm c \int d^4x \left(A_\nu^{R/L}F_{\rho\sigma}^{R/L}-A_\nu^{L/R}F_{\rho\sigma}^{R/L}+2A_\nu^{R/L}F_{\rho\sigma}^{L/R}\right)
\delta A_\mu^{L/R} \epsilon^{\mu\nu\rho\sigma} \non
&& \mp \,2c\int d^3x\, A_\nu^{R/L}A_\sigma^{L/R}\delta A_\mu^{L/R}\Big|_{x_\rho} 
\epsilon^{\mu\nu\rho\sigma} \, .
\eea
Again dropping the space-time surface terms, the contribution to the currents is
therefore
\be \label{DeltaJ}
\Delta {\cal J}^\mu_{L/R} = \mp c \left(A_\nu^{R/L}F_{\rho\sigma}^{R/L}-A_\nu^{L/R}F_{\rho\sigma}^{R/L}+2A_\nu^{R/L}F_{\rho\sigma}^{L/R}\right)
\epsilon^{\mu\nu\rho\sigma} \, ,
\ee
and the contribution to the divergence of the currents becomes 
\be
\partial_\mu \Delta {\cal J}^\mu_{L/R} = 
\mp c\left(F_{\mu\nu}^{R/L}\widetilde{F}^{\mu\nu}_{R/L} +F_{\mu\nu}^{L/R}\widetilde{F}^{\mu\nu}_{R/L}\right) \, .
\ee
Denoting renormalized left- and right-handed currents as
\be \label{tildeJ}
\bar{\cal J}^\mu_{L/R} \equiv {\cal J}^\mu_{L/R} + \Delta{\cal J}^\mu_{L/R} \, , 
\ee
and similarly the renormalized axial and vector currents as
$\bar{\cal J}_\mu$, $\bar{\cal J}_\mu^5$, we find that the choice 
\be \label{c}
c=\frac{N_c}{48\pi^2} 
\ee
leads to the {\em covariant} anomaly
\begin{subequations}\label{covanomaly}
\bea
\partial_\mu \bar{\cal J}^\mu &=& 0  \, , \label{covanomaly1} \\
\partial_\mu \bar{\cal J}^\mu_5 &=& \frac{N_c}{8\pi^2} F_{\mu\nu}^V \widetilde{F}^{\mu\nu}_V
+ \frac{N_c}{24\pi^2} F_{\mu\nu}^A \widetilde{F}^{\mu\nu}_A \, . \label{covanomaly2} 
\eea
\end{subequations}
Note that the prefactor in front of the first term in the axial anomaly now has changed to $N_c/(8\pi^2)$, from $N_c/(24\pi^2)$ in 
eq.\ (\ref{consanomaly1b}), which is the well-known result for the
Adler-Bell-Jackiw anomaly for QED \cite{Bell:1969ts,Adler:1969gk}
and which is essential for getting the correct pion decay rate $\pi^0\to 2\gamma$. 
The necessity of adding the counterterm (\ref{DeltaS}) to the Sakai-Sugimoto
model is in fact completely analogous to the very same and well-known procedure 
in chiral models where a Wess-Zumino-Witten term accounts for the anomaly
\cite{Kaymakcalan:1983qq}. 

In the literature sometimes the coefficient of the subleading term in the asymptotic
behavior of $A_\mu(x,|z|\to\infty)$ and thus the YM part of the current
(see eq.~(\ref{expandYM}))
is identified with the full current \cite{Yee:2009vw},
see also \cite{Erdmenger:2008rm,Banerjee:2008th,Torabian:2009qk,Son:2009tf}. Using this
identification, it has also been assumed that the equation of motion for $A_z$ (\ref{eq2}) represents the anomaly equation \cite{Lifschytz:2009si}.
Indeed, from eq.\ (\ref{eq2}) one obtains the apparent anomaly 
\be \label{anomalywrong}
\partial_\mu {\cal J}_{{\rm YM},L/R}^\mu = \mp \frac{N_c}{16\pi^2} F_{\mu\nu}^{L/R}\widetilde{F}_{L/R}^{\mu\nu} \, ,
\ee
which leads to
\begin{subequations}\label{YMcurranomalies}
\bea
\partial_\mu{\cal J}^\mu_{\rm YM} &=& \frac{N_c}{4\pi^2} F_{\mu\nu}^V\widetilde{F}^{\mu\nu}_A \, , \\
\partial_\mu{\cal J}^\mu_{\rm YM,5} &=& \frac{N_c}{8\pi^2} \left(F_{\mu\nu}^V\widetilde{F}^{\mu\nu}_V+F_{\mu\nu}^A\widetilde{F}^{\mu\nu}_A\right) \, ,
\eea
\end{subequations}
and this does contain the same coefficient in front of $F_{\mu\nu}^V\widetilde{F}^{\mu\nu}_V$ as the full covariant anomaly (\ref{covanomaly}).
However, it differs from the latter
in the presence of axial gauge fields. In particular, the vector current is then
not strictly conserved. 
The renormalized current
$\bar{\mathcal J}_{L/R}$ satisfies 
eq.\ (\ref{anomalywrong})
only for $F_{\mu\nu}^L=F_{\mu\nu}^R\,$.%
\footnote{The more general validity of eq.\ (\ref{anomalywrong})
has been assumed incorrectly in eq.\ (2.1) of ref.\ \cite{Yee:2009vw}
and eq.\ (36) of ref.\ \cite{Fukushima:2008xe}.} 
Even when this issue may be ignored, because all
axial vector field strengths are set to zero, it appears to be questionable
to keep only part of the full current (\ref{currents}).

In the 
remainder of the paper we shall consider the full currents for which Bardeen's counterterm is needed, and study the implications, 
which indeed differ from keeping only the YM part of the currents. (The effect of truncating to the YM part can be
easily read off from the expressions that we shall give.)

\section{Background electromagnetic fields and chemical potentials}
\label{sec:background}

The discussion in the previous section was general in the sense that we have not specified any gauge fields except for the gauge choice
$A_z=0$. In this section we specify our ansatz according to the physical situation we are interested in. This includes a background
magnetic field $B$ as well as separate left- and right-handed chemical potentials
$\mu_{L,R}=A_0(z=\pm\infty)$, or, equivalently, the ordinary
quark chemical potential $\mu=(\mu_R+\mu_L)/2$ and 
an axial chemical potential $\mu_5=(\mu_R-\mu_L)/2$. 
With these ingredients we can obtain
results relevant for the heavy-ion context 
(nonzero $\mu_5$, negligibly small $\mu$) and for the astrophysical context 
(vanishing $\mu_5$, large $\mu$). In order to be able
to check the axial anomaly explicitly, we also add an electric field $E$ and an ``axial electric field'' $\epsilon$ 
parallel to the magnetic field. The electric field $E$ is needed because
the axial anomaly is proportional to $\mathbf E\cdot \mathbf B$. 
The (unphysical) field $\epsilon$ shall be used to check the absence of
a vector anomaly, i.e., the conservation of the vector current, which must be true even in the presence of $\epsilon$, and would 
be trivial without $\epsilon$. In our final results for the
currents, the electric fields
are however set to zero.

For previous discussions of background electric and magnetic fields in the Sakai-Sugimoto model
see for instance refs.\ \cite{Bergman:2008qv,Thompson:2008qw,Rebhan:2008ur,Bergman:2008sg,Johnson:2008vna}. We shall only consider
spatially homogeneous systems. This is the simplest case, 
which might however require generalization when the true ground state is
more complicated, for instance when Skyrme crystals are formed
\cite{Kim:2007vd}.

\subsection{Chirally broken phase}
\label{sec:solution1}

In our ansatz the nonzero fields are $A_0(t,z)$, $A_1(x_2)$, $A_3(t,z)$,
where the dependence on $t$ will only be present for nonvanishing electric
fields $E$ and $\epsilon$ at the holographic boundary.
The temporal component $A_0$ is needed to account for nonzero (left- and right-handed) chemical potentials which correspond to the 
values of $A_0$ at the boundary. The electromagnetic fields are encoded in the boundary values of the spatial components.
Since the gauge symmetry in the bulk corresponds to a global
symmetry for the dual field theory, the fields at the holographic boundary
are not dynamical and merely serve as background fields.
This is however sufficient for our purpose. 
The magnetic field ${\bf B}$ is assumed to point into the 3-direction, ${\bf B}=(0,0,B)$. Consequently, we can choose 
\be
A_1(x_2) = - x_2 B   
\ee
at the holographic boundary. The equations of motion show that $A_1$ can be chosen to be constant in $z$ throughout the bulk. (This is 
different in the presence of an isospin chemical potential \cite{Rebhan:2008ur}.) Consequently, $F_{12}(x,z)=B$. For notational convenience 
we have absorbed the electric quark charge $q_f$ into $B$, i.e., actually $B\to q_f B$ with $q_f = 2/3\,e$ for $f=u$, and $q_f = -1/3\,e$ for $f=d$. 
With nonzero $A_0$ and $A_1$, accounting for the chemical potential and the magnetic field, a nonzero $A_3$ is induced, even without 
electric field. In the broken phase, $A_3$ develops a nonzero boundary value, corresponding to the gradient of the neutral pion 
\cite{Bergman:2008qv,Thompson:2008qw,Rebhan:2008ur}. Just as for a usual superfluid, where the gradient of the phase of the
order parameter is proportional to the superfluid velocity, this gradient of the pion field can be viewed as an axial
supercurrent \cite{Rebhan:2008ur}.
 
Next we introduce the electric field ${\bf E}=(0,0,E)$ parallel to ${\bf B}$ and, as explained above, an ``axial electric field'' 
$\bm{\epsilon}=(0,0,\epsilon)$. We thus have to add $-t(E\mp\epsilon)$ to the boundary 
value of $A_3$, such that, together with the axial supercurrent $\jmath_t$, we have 
\be \label{A3inf}
A_3(t,z=\pm\infty) = -t(E\mp\epsilon) \mp \jmath_t \, .
\ee
Due to the axial electric field we allow the supercurrent to become time-dependent.
Strictly speaking the electric fields prevent us from using a thermodynamic description 
since it introduces a time-dependence and thus non-equilibrium physics. Therefore, our electric field should be considered infinitesimal. 
This is sufficient for our purpose since we can check the anomaly relations with an arbitrarily small electric field. Moreover, as mentioned
above, the physical situations we are interested in do not require finite electric fields anyway. 

With the above ansatz, the YM and CS contributions to the action (\ref{SYMCS}) become 
\begin{subequations}\label{action}
\bea 
S_{\rm YM} &=& \kappa M_{\rm KK}^2 \int d^4x \int_{-\infty}^{\infty} dz\,k(z)\left[-(\partial_z A_0)^2+(\partial_z A_3)^2\right] \, , 
\label{actionYM}\\
S_{\rm CS} &=& \frac{N_c}{12\pi^2}  \int d^4x \int_{-\infty}^{\infty} dz\,
\Big\{(\partial_2 A_1)\left[A_0(\partial_z A_3)-A_3(\partial_z A_0)\right] \non
&& \hspace{2.5cm} -\,A_1 \left[(\partial_2 A_0)(\partial_z A_3)-(\partial_2 A_3)(\partial_z A_0)\right]\Big\} 
\, . \label{actionCS}
\eea
\end{subequations}
We have written all terms which are needed to derive the equations of motion, including the ones that vanish on-shell. 
More specifically, the second line in the CS action (\ref{actionCS}) vanishes on-shell because neither $A_0$ nor $A_3$ depends on $x_2$, 
but yields a finite contribution to the equations of motion. The equations of motion are
\begin{subequations} \label{E123}
\bea
\partial_z(k\partial_z A_0) &=& 2\beta \partial_z A_3 \, , \label{E1} \\
\partial_z(k\partial_z A_3) &=& 2\beta \partial_z A_0 \, , \label{E2} \\
\partial_t(k\partial_z A_0) &=& 2\beta \partial_t A_3 \, , \label{E3} 
\eea
\end{subequations}
with the dimensionless magnetic field  
\be \label{defbetabroken}
\beta\equiv \frac{\alpha B}{M_{\rm KK}^2} \, ,
\ee
and $\alpha\equiv 27\pi/(2\lambda)$. We defer the details of solving the equations of motion to appendix \ref{app:broken}.
The results for the gauge fields and field strengths are  
\begin{subequations} \label{gaugefields1}
\bea
A_0(t,z) &=& \mu_t-\mu_{5,t}\frac{\sinh(2\beta\arctan z)}{\sinh\beta\pi} \non
&&-(\jmath_t-\epsilon t) \left[\frac{\cosh(2\beta\arctan z)}{\sinh\beta\pi} -\coth\beta\pi\right] \, , \\
A_3(t,z) &=& -tE-\mu_{5,t}\left[\frac{\cosh(2\beta\arctan z)}{\sinh\beta\pi}-\coth\beta\pi\right] \non
&&-(\jmath_t-\epsilon t) \frac{\sinh(2\beta\arctan z)}{\sinh\beta\pi}  \, ,
\eea
\end{subequations}
and
\begin{subequations} \label{fieldstrengths1}
\bea
k\partial_z A_0 &=& -2\beta\left[\mu_{5,t}\frac{\cosh(2\beta\arctan z)}{\sinh\beta\pi}
+(\jmath_t-\epsilon t)\frac{\sinh(2\beta\arctan z)}{\sinh\beta\pi}\right] \, , \\
k\partial_z A_3 &=& -2\beta\left[\mu_{5,t}\frac{\sinh(2\beta\arctan z)}{\sinh\beta\pi}
+(\jmath_t-\epsilon t)\frac{\cosh(2\beta\arctan z)}{\sinh\beta\pi}\right] \, .
\eea
\end{subequations}
Here we have denoted  
\be\label{mutbroken}
\mu_{t} \equiv \mu + \epsilon t\coth\beta\pi \, ,  \qquad \mu_{5,t} \equiv \mu_5 + Et\tanh\beta\pi \, , 
\ee
i.e., both boundary values of $A_0$ become time-dependent through the electric fields. As can be seen from the detailed 
derivation in appendix \ref{app:broken}, this time-dependence is unavoidable.  
In fig.\ \ref{figgauge1} we plot the gauge fields for $E=\epsilon=0$ 
at the minimum, i.e., after $\jmath_t$ has been determined to minimize the free energy, see below. 

\FIGURE[t]{\label{figgauge1}
{ \centerline{\def\epsfsize#1#2{0.7#1}
\epsfbox{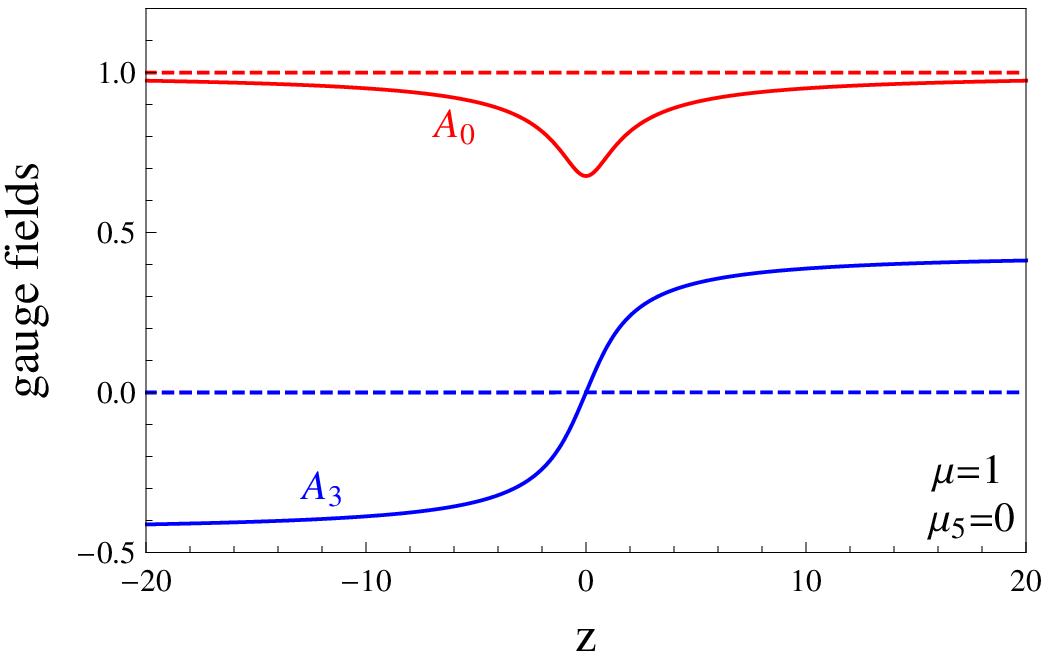}
\epsfbox{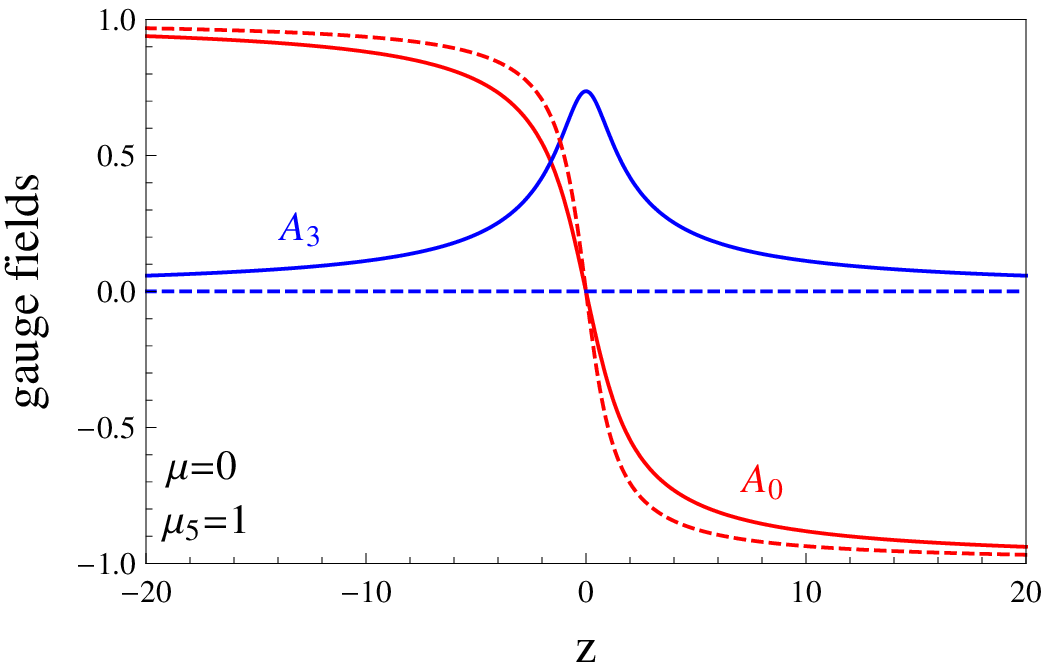}}
}
\caption{Gauge fields in the chirally broken phase as functions of the holographic coordinate $z\in[-\infty,\infty]$ 
for a finite quark chemical potential and vanishing axial chemical potential 
(left) and vice versa (right). Dashed lines: gauge fields with vanishing magnetic field; solid lines: gauge fields with 
a nonzero magnetic field $\beta=0.6$. In both plots we have set the electric fields to zero, $E=\epsilon=0$. 
The boundary values at $z=\pm \infty$ correspond to left- and right-handed quantities. The magnetic field 
induces an axial supercurrent (boundary value of $A_3$) in the case of a nonvanishing quark chemical potential. If both
$\mu$ and $\mu_5$ are nonvanishing, the gauge fields are neither symmetric nor antisymmetric in $z$. The analytic expressions
for these curves are given in eqs.\ (\ref{gaugefields1}).
 }  
}

The thermodynamic potential $\Omega=\frac{T}{V}S_{\rm on-shell}$ is obtained from eqs.\ (\ref{YMCSonshell}), treating $t$ as an external parameter,
\be \label{Omega1}
\Omega = \frac{8\kappa M_{\rm KK}^2}{3}\Big\{\left[(\jmath_t-\epsilon t)^2-\mu_{5,t}^2\right]\rho(\beta)+\beta\left[\mu_t(\jmath_t-\epsilon t) 
+tE\mu_{5,t}\right] \Big\} \, ,
\ee
where we have abbreviated
\be \label{defrho}
\rho(\beta)\equiv \beta\coth\beta\pi+\frac{\pi\beta^2}{2\sinh^2\beta\pi}\simeq \left\{\begin{array}{cc} 
\displaystyle{\frac{3}{2\pi} +\frac{\pi}{6}\beta^2}  &\;\;
{\rm for} \; \beta\to 0 \\[2ex] |\beta| & {\rm for}\;|\beta| \to \infty \end{array}\right.\, .
\ee
Minimization of $\Omega$ with respect to $\jmath_t$ yields the axial supercurrent
\be \label{jmin}
\jmath_t = - \frac{\beta \mu}{2\rho(\beta)}+\epsilon t\left[1-\frac{\beta\coth\beta\pi}{2\rho(\beta)}\right] \, .
\ee
We see that the supercurrent depends neither on $\mu_5$ nor on $E$. Therefore, at $t=0$ it is simply the one-flavor limit of the 
result obtained in ref.\ \cite{Rebhan:2008ur} (where the D8 and $\overline{\rm D8}$ branes were identified with $R$ and $L$, not with 
$L$ and $R$, respectively, hence the different sign of the supercurrent). 


\subsection{Chirally symmetric phase}
\label{sec:solution2}

As explained in sec.\ \ref{sec:brief}, in the chirally symmetric phase the D8 and $\overline{\rm D8}$-branes are not connected. On both 
branes the holographic coordinate $z$ now
runs from $z=0$, the black hole horizon, to the holographic boundary
at $z=\infty$, and both branes yield separate contributions to the action,
\be\label{SRL}
S = (S_{\rm YM}^L + S_{\rm YM}^R) + (S_{\rm CS}^L - S_{\rm CS}^R) \, .
\ee
The CS action assumes different overall signs on the D8- and $\overline{\rm D8}$-branes since its parity is odd.
The YM and CS contributions are  
\begin{subequations}\label{actionsym}
\bea 
S_{\rm YM}^h &=& \kappa M_{\rm KK}^2 \theta^3 \int d^4x \int_{0}^{\infty} dz\,\left[-k_0(z)(\partial_z A_0^h)^2
+k_3(z)(\partial_z A_3^h)^2\right] 
\, , 
\label{actionYMsym}\\
S_{\rm CS}^h &=& \frac{N_c}{12\pi^2}  \int d^4x \int_{0}^{\infty} dz\,
\Big\{(\partial_2 A_1^h)[A_0^h(\partial_z A_3^h)-A_3^h(\partial_z A_0^h)] \non
&& \hspace{1.5cm} -\,A_1^h [(\partial_2 A_0^h)(\partial_z A_3^h)-(\partial_2 A_3^h)
(\partial_z A_0^h)]\Big\} 
\, , \label{actionCSsym}
\eea
\end{subequations}
with $h=L,R$. Here, we have defined the dimensionless temperature
\be
\theta \equiv \frac{2\pi T}{M_{\rm KK}} \, . 
\ee
In contrast to the broken phase there are different metric functions for temporal and spatial components of the gauge fields,  
\be \label{defk}
k_0(z)\equiv \frac{(1+z^2)^{3/2}}{z} \, , \qquad k_3(z)\equiv z(1+z^2)^{1/2} \, .
\ee
Note the slight difference in notation of the gauge fields: while in the broken phase $A_\mu^{L/R}(x)\equiv A_\mu(x,z=\pm\infty)$ 
always implies evaluation at the holographic boundary, here we label the bulk gauge fields $A_\mu^{L/R}(x,z)$ by $L$ and $R$
to indicate whether they live on the D8- or on the $\overline{\rm D8}$-brane. Since we always discuss broken and symmetric phases separately, 
this should not cause any confusion. 
 
The equations of motion on the separate branes become 
\begin{subequations} \label{E123sym}
\bea
\partial_z(k_0\partial_z A_0^{L/R}) &=& \pm \frac{2\beta}{\theta^3} \partial_z A_3^{L/R} \, , \label{E1sym} \\
\partial_z(k_3\partial_z A_3^{L/R}) &=& \pm \frac{2\beta}{\theta^3} \partial_z A_0^{L/R} \, , \label{E2sym} \\
\partial_t(k_0\partial_z A_0^{L/R}) &=& \pm \frac{2\beta}{\theta^3} \partial_t A_3^{L/R} \, . \label{E3sym} 
\eea
\end{subequations}
Details of solving the equations of motion are presented in appendix \ref{app:restored}.
The final solution for the gauge fields is
\begin{subequations} \label{gaugefields2}
\bea
A_0^{L/R}(t,z) &=& (\mu_t\mp\mu_{5,t})\left[p(z)-\frac{p_0}{q_0}\,q(z)\right] \, , \\
A_3^{L/R}(t,z) &=& -t(E\mp \epsilon)\pm\frac{\mu_t\mp\mu_{5,t}}{2\beta/\theta^3}
\left[k_0\partial_zp-\frac{p_0}{q_0}(1+k_0\partial_zq)\right]
\, ,
\eea
which is plotted in fig.\ \ref{figgauge2} for $E=\epsilon=0$.
\end{subequations}
Below we shall also need the field strengths on the branes, 
\begin{subequations} \label{fieldstrengths2}
\bea
k_0\partial_z A_0^{L/R} &=&(\mu_t\mp\mu_{5,t})
\left(k_0\partial_zp-\frac{p_0}{q_0}k_0\partial_zq\right) 
\, , \\
k_3\partial_z A_3^{L/R} &=& \pm \frac{2\beta}{\theta^3} (\mu_t\mp\mu_{5,t})\left[p(z)-\frac{p_0}{q_0}q(z)\right] \, .
\eea
\end{subequations}
\FIGURE[t]{\label{figgauge2}
{ \centerline{\def\epsfsize#1#2{0.7#1}
\epsfbox{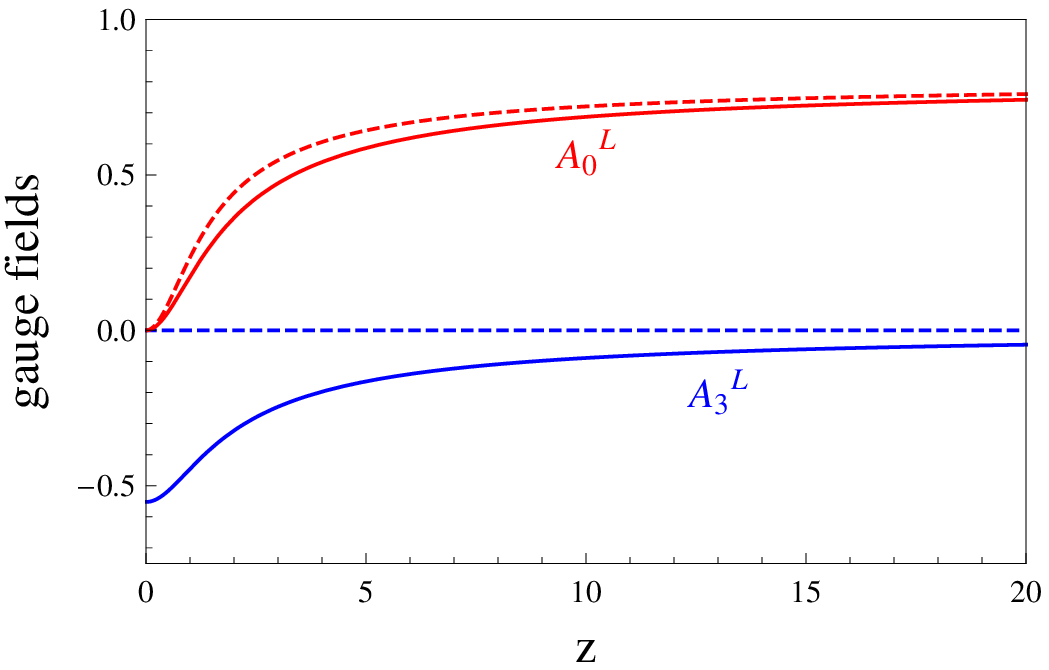}
\epsfbox{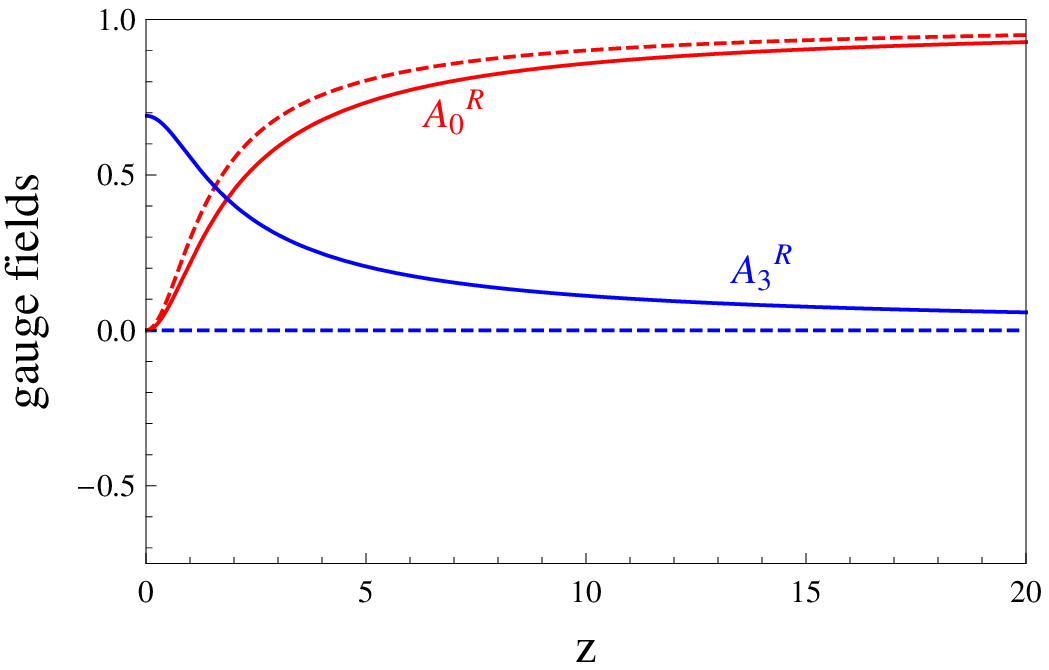}}
}
\caption{ 
Left- and right-handed gauge fields (left and right panel, respectively) in the chirally symmetric phase  
 as functions of the holographic coordinate $z\in [0,\infty]$ 
for $\mu=0.9$, $\mu_5=0.1$. We have set the electric fields to zero, $E=\epsilon=0$.
The temporal components $A_0^{L/R}$ approach the chemical 
potentials $\mu\mp\mu_5$ at the boundary $z=\infty$, while the spatial components $A_3^{L/R}$ vanish at $z=\infty$. A finite magnetic field 
(solid lines, here $\beta/\theta^3=0.6$) 
distorts the gauge fields compared to the case of vanishing magnetic fields (dashed lines). In particular, 
the spatial component develops a nonzero value at $z=0$. The different sign of this value for left- and right-handed fields, i.e., on the
D8- and $\overline{\rm D8}$-branes, ensures the correct parity behavior of the fields.
The analytical expressions for these curves are given in eqs. (\ref{gaugefields2}).
 }  
}
The functions $p(z)$, $q(z)$ are hypergeometric functions which we defined in eqs.\ (\ref{pq}) and which depend on the ratio $\beta/\theta^3$. 
Their values at $z=0$ are denoted by $p_0$, $q_0$, see eqs.\ (\ref{tauzero1}), and the ratio $p_0/q_0$ behaves for small and large magnetic fields
as 
\be \label{limitstau}
\frac{p_0}{q_0} \simeq \left\{\begin{array}{cc} 1+(2\beta/\theta^3)^2(\ln 4-1) &\;\;
{\rm for} \; \beta/\theta^3\to 0 \\[2ex] 2|\beta|/\theta^3 & {\rm for}\;|\beta|/\theta^3 \to \infty \end{array}\right. \, . 
\ee
The boundary values of the temporal components are
$A_0^{L/R}(t,z=\infty)=\mu_t\mp\mu_{5,t}$ with  
\be \label{mut1}
\mu_t \equiv \mu+2t\epsilon\frac{\beta}{\theta^3}\frac{q_0}{p_0} \, , \qquad
\mu_{5,t} \equiv \mu_5+2 tE\frac{\beta}{\theta^3}\frac{q_0}{p_0} \, .
\ee
It is instructive to compare this behavior of the axial chemical potential with the expected behavior for free fermions
in a magnetic field. To this end, consider the lowest Landau level in which the spin of all (say, positively charged) fermions is aligned parallel
to the magnetic field. As a consequence, all right- (left-) handed massless fermions move parallel (antiparallel) to the magnetic field. An electric
field parallel to the magnetic field now shifts all momenta in the positive 3-direction by an amount $Et$. Consequently, some of the left- 
handed fermions are converted into right-handed fermions and a shift $Et$ is induced in the difference 
of right- and left-handed Fermi momenta, $(p_F^R-p_F^L)/2 = Et$ \cite{Fukushima:2008xe,Ambjorn:1983hp}. 
Interpreting $\mu_{5,t}$ as $(p_F^R-p_F^L)/2$ (strictly speaking there is no well-defined Fermi momentum in our model),
eq.\ (\ref{mut1}) reproduces this shift for asymptotically large magnetic fields because in this case $q_0/p_0\to \theta^3/(2\beta)$.
For small magnetic fields $q_0/p_0\to 1$, and the shift becomes linear in the magnetic field. Since $\beta/\theta^3 \propto B/T^3$, 
we can in principle also obtain $\mu_{5,t} = \mu_5+ tE$ for sufficiently small 
temperatures and fixed magnetic field. However, we cannot reduce the temperature arbitrarily in the above expression 
since below the critical temperature $T_c$ we are in the chirally broken phase. 
In this case the analogous, temperature-independent relation in eq.\ (\ref{mutbroken}) holds.

The free energy, obtained from the YM and CS contributions (\ref{YMCSOmega2}), is 
\be \label{Omega2}
\Omega = -\frac{2\kappa M_{\rm KK}^2}{3}\left[\theta^3(\mu_t^2+\mu_{5,t}^2)\eta-4\beta\,t\left(\mu_t\epsilon+\mu_{5,t}E\right)\right] \, ,
\ee
where we introduced the function
\be \label{defeta}
\eta(\beta/\theta^3) \equiv I_0-(2\beta/\theta^3)^2 I_3
+2\frac{p_0}{q_0} \simeq \left\{\begin{array}{cc} 3+(2\beta/\theta^3)^2(\ln 4-1) &\;\;
{\rm for} \; \beta/\theta^3 \to 0 \\[1ex] 4|\beta|/\theta^3 & {\rm for}\; |\beta|/\theta^3 \to \infty \end{array}\right. \, , 
\ee
with integrals $I_0$ and $I_3$ defined in eqs.\ (\ref{defI}).

\subsection{Ambiguity of currents}
\label{sec:currents}

In the following discussion we restrict ourselves to the symmetric phase, but one can easily check that all arguments hold for the broken 
phase as well. Let us first give the analogue of the definition of the currents (\ref{currents}) for the symmetric phase,
\be \label{currentssym}
{\cal J}_{L/R}^\mu = -\left(2\kappa M_{\rm KK}^2\theta^3 k_{(\mu)} F_{L/R}^{z\mu}\mp
\frac{N_c}{24\pi^2}\epsilon^{\mu\nu\rho\sigma}A_\nu^{L/R}F_{\rho\sigma}^{L/R}\right)_{z=\infty} \, ,
\ee
where the notation $k_{(\mu)}$ (no summation over $\mu$) indicates the different metric functions for temporal and 
spatial components, see eq.\ (\ref{defk}). Equivalently, and in analogy to eq.\ (\ref{dLdA}), we can write the currents in the symmetric phase
as 
\be \label{currsym}
{\cal J}_{L/R}^\mu = \left. -\frac{\partial {\cal L}}{\partial\,\partial_zA_\mu^{L/R}}\right|_{z=\infty} \, . 
\ee
We shall now show that the currents defined via these equations are different from the ones obtained via taking the derivative of the
free energy (\ref{Omega2}) with respect to the corresponding source. We do so for the vector density, i.e., the sum of 
left- and right--handed 0-components of the currents. One can observe the same ambiguity for the other nonvanishing components.
The following arguments do not depend on the electric fields, so we temporarily set $\epsilon=E=0$ for simplicity 
(and for a truly equilibrated situation).
From the definition (\ref{currentssym}) and the gauge fields (\ref{gaugefields2}) and field strengths (\ref{fieldstrengths2})
we obtain 
\be \label{J01}
{\cal J}^0 = {\cal J}^0_R + {\cal J}^0_L = 4\kappa M_{\rm KK}^2 \theta^3  \frac{p_0}{q_0}\mu \, .
\ee
On the other hand, the free energy $\Omega$ of the system should yield the number density via the thermodynamic relation
\be \label{thermo}
n = -\frac{\partial\Omega}{\partial\mu} = \frac{4\kappa M_{\rm KK}^2\theta^3}{3}\mu\,\eta \, . 
\ee
This result shows that $n\neq {\cal J}^0$ which, given spatial homogeneity, is inconsistent. This inconsistency is absent for 
vanishing magnetic fields: using the behavior of the functions $p_0/q_0$ and $\eta$ from eqs.\ (\ref{limitstau}) and 
(\ref{defeta}) one sees that for $\beta=0$ the expressions for ${\cal J}^0$ and $n$ are identical. We can formulate this 
observation in a more general way. To this end we write the left- and right-handed on-shell Lagrangians, i.e., the integrands of 
the on-shell action (\ref{SRL}), as ${\cal L}_h(A_0^h,\partial_zA_0^h,A_3^h,\partial_z A_3^h)$, where all arguments 
of ${\cal L}_h$ depend on the chemical potentials $\mu_h$ with $h=L,R$ and $\mu_{L/R}=\mu\mp\mu_5$. Then, with 
$\Omega_h=T/V\,\int d^4x\int_0^\infty dz\, {\cal L}_h$ we have  
\bea
\frac{\partial\Omega_h}{\partial\mu_h} &=& 
\frac{T}{V}\sum_{i=0,3}\int d^4x \int_0^\infty dz\,\left(\frac{\partial{\cal L}_h}{\partial A_i^h}\frac{\partial A_i^h}{\partial\mu_h}
+\frac{\partial{\cal L}_h}{\partial \, \partial_zA_i^h}\frac{\partial \,\partial_z A_i^h}{\partial\mu_h}\right) \non
&=& \frac{T}{V}\sum_{i=0,3}\left[
\int d^4x \left.\frac{\partial{\cal L}_h}{\partial \, \partial_zA_i^h}\frac{\partial A_i^h}{\partial\mu_h}\right|_{z=0}^{z=\infty}
+\int d^4x \int_0^\infty dz\,
\partial_2\frac{\partial{\cal L}_h}{\partial\,\partial_2 A_i^h}\frac{\partial A_i^h}{\partial\mu_h}\right] \, ,
\eea
where we have used partial integration and added and subtracted the derivative term in $x_2$ in order to make use of the equations of motion.
Now we use 
\be
\left.\frac{\partial A_0^h}{\partial\mu_h}\right|_{z=\infty} = 1\, , \qquad \left.\frac{\partial A_0^h}{\partial\mu_h}\right|_{z=0}=
\left.\frac{\partial A_3^h}{\partial\mu_h}\right|_{z=\infty} = \left.\frac{\partial {\cal L}_h}{\partial\,\partial_zA_3^h}\right|_{z=0} = 0\, , 
\ee
which follows from the explicit solutions (\ref{gaugefields2}), whose behavior at $z=0$, $z=\infty$ is obtained with the 
help of eqs.\ (\ref{tauinfty}), (\ref{tauzero1}), and (\ref{tauzero2}). With these relations and the definition of the current from 
eq.\ (\ref{currsym}) we obtain
\be \label{dOdmu} 
-\frac{\partial\Omega}{\partial\mu} = {\cal J}^0 -
\frac{T}{V}\sum_{h=L,R}\sum_{i=0,3}\int d^4x \int_0^\infty dz\,\partial_2\frac{\partial{\cal L}_h}{\partial\,\partial_2 A_i^h}
\frac{\partial A_i^h}{\partial\mu_h} \, .
\ee
This is the general form of the difference between the density defined as the 0-component of the current defined via eq.\ (\ref{currsym})
and the density defined via the thermodynamic relation (\ref{thermo}). For an explicit check of this relation one  
inserts the expressions
\begin{subequations}
\bea
\partial_2\frac{\partial{\cal L}_{L/R}}{\partial\,\partial_2 A_0^{L/R}} &=& \pm\frac{4\kappa M_{\rm KK}^2}{3}\beta\partial_z A_3^{L/R} \, , 
\\
\partial_2\frac{\partial{\cal L}_{L/R}}{\partial\,\partial_2 A_3^{L/R}} &=& \mp\frac{4\kappa M_{\rm KK}^2}{3}\beta\partial_z A_0^{L/R} \, , 
\eea
\end{subequations}
and 
\begin{subequations}
\bea
\frac{\partial A_0^{L/R}}{\partial\mu_{L/R}} &=& p(z)-\frac{p_0}{q_0}q(z) \, , \\
\frac{\partial A_3^{L/R}}{\partial\mu_{L/R}} &=& \pm\frac{\theta^3}{2\beta}\left[k_0\partial_z p-\frac{p_0}{q_0}
\left(1+k_0\partial_zq\right)\right] \, , 
\eea
\end{subequations}
into eq.\ (\ref{dOdmu}). This yields 
\be
-\frac{\partial\Omega}{\partial\mu}={\cal J}^0 + \frac{4\kappa M_{\rm KK}^2\theta^3}{3}\mu \left[I_0-\left(\frac{2\beta}{\theta^3}\right)^2 I_3 
-\frac{p_0}{q_0}\right] \, . 
\ee
With the definition (\ref{defeta}) this confirms the difference between $n$ and ${\cal J}^0$ obtained from eqs.\ (\ref{J01})
and (\ref{thermo}). 

From the general form (\ref{dOdmu}) we see that the additional term is a boundary term at the spatial boundary of the system. 
This suggests that the ambiguity in the currents is related to the terms we have dropped in sec.\ \ref{sec:action}, 
see eqs.\ (\ref{currboundary}). These terms correspond to currents at the
spatial boundary and disappear in the presence of a homogeneous magnetic field 
only if the variation $\delta A_\mu(x,z=\pm\infty)$ can be chosen to vanish at this boundary.
So this problem might be 
resolved by considering more complicated, spatially inhomogeneous gauge fields. In our homogeneous ansatz,  
it is however
a priori not clear which definition of the currents corresponds to the correct physics. 

A possible solution to this ambiguity was suggested and applied in 
refs.\ \cite{Bergman:2008qv,Lifschytz:2009si,Lifschytz:2009sz}. In these references, 
the CS action has been modified according to 
\bea \label{Sprime}
S_{\rm CS}^{\prime\, h} &=& \frac{N_c}{12\pi^2}\int d^4x\int_0^\infty dz\,
\left\{\frac{3}{2}(\partial_2 A_1^h)\left[A_0^h(\partial_z A_3^h)-A_3^h(\partial_z A_0^h)\right]
\right.\non
&&\left. \hspace{2.5cm}-\,\frac{1}{2}(\partial_z A_1^h)\left[A_0^h(\partial_2 A_3^h)-A_3^h(\partial_2 A_0^h)\right] \right\} \, .
\eea
This modified action (marked by a prime) 
is obtained from the original CS action (\ref{actionCSsym}) by adding a boundary term at the holographic and the 
spatial boundary,
\be
S_{\rm CS}^{\prime\, h} = S_{\rm CS}^h +  S_{\rm boundary}^h\, , 
\ee
with 
\bea \label{Sboundary}
S_{\rm boundary}^h &=& \frac{N_c}{24\pi^2}\left\{\int d^3x \int_0^\infty dz\, A_1^h\Big[A_0^h(\partial_zA_3^h)-A_3^h(\partial_zA_0^h)\Big]_{x_2}
\right. \non
&&\left.\hspace{1cm}-\,\int d^4x \, A_1^h\Big[A_0^h(\partial_2A_3^h)-A_3^h(\partial_2A_0^h)\Big]_{z=0}^{z=\infty}\right\} \, .
\eea
Note that this boundary term cannot be considered as a holographic counterterm
since it involves an integration over $z$. 
From eq.\ (\ref{Sprime}) we see that the addition
of $S_{\rm boundary}^h$ 
effectively amounts to a multiplication of the on-shell action by 3/2 because 
the second line in eq.\ (\ref{Sprime}) vanishes on-shell. 
The benefit of the modified action is that the integrand on the right-hand side of eq.\ (\ref{dOdmu}) 
vanishes now, i.e., there is no ambiguity in the currents anymore. 

Modifications of a CS action by boundary terms are in fact sometimes
necessary in order to ensure validity of the variational principle
in the presence of nontrivial boundary values \cite{Elitzur:1989nr}.
However, this is not what the above modification is achieving. Instead,
it leads to gauge invariance under the residual gauge transformations $A_1\to A_1+\partial_1 \Lambda(x_1)$ which are compatible with the boundary conditions
of our ansatz and
which do not vanish at spatial infinity $x_2=\pm\infty$ \cite{Bergman:2008qv}. 
In fact, by this modification one loses all anomalies for the (now uniquely defined) currents, as we show now. 
To this end, we switch on the electric fields again. Then, the currents of the original action in the symmetric phase are 
\begin{subequations} \label{J0123sym}
\bea
T>T_c: \quad {\cal J}^0_{L/R} &=& 2\kappa  M_{\rm KK}^2\frac{p_0}{q_0}\left[\theta^3(\mu_t\mp\mu_{5,t})\pm\frac{2}{3}\frac{q_0}{p_0}
\beta (E\mp\epsilon)t\right] \, , \\
{\cal J}^1_{L/R} &=& 0 \, , \\
{\cal J}^2_{L/R} &=& \pm \frac{4\kappa  M_{\rm KK}^2}{3}\beta \,x_2 (E\mp\epsilon) \, , \label{J2sym}\\
{\cal J}^3_{L/R} &=& \mp 4\kappa  M_{\rm KK}^2\beta\left(\mu_t \mp \mu_{5,t}\right)\left(1-\frac{1}{3}\right) \, , \label{J3sym}
\eea
\end{subequations}
where we have used the definition (\ref{currentssym}) and the gauge fields (\ref{gaugefields2}) and field strengths (\ref{fieldstrengths2}).
All terms containing a 1/3 originate from the CS contribution of the current, i.e., from the second term in eq.\ (\ref{currentssym}). 
All other terms are YM contributions. In particular,
the 2-component of the current is a pure CS term. This component is unphysical because it depends on our choice to 
introduce the magnetic field via the gauge field $A_1$. We could have introduced the same magnetic field via $A_2$ or a 
combination of $A_1$ and $A_2$, in which case the 1- and 2-components of the currents would have been different. We shall see below that 
Bardeen's counterterm solves this problem by canceling the 2-component. Here, however, it gives a nonzero contribution to the 
anomaly. Namely, the divergence of the (unmodified) currents becomes
\be
\partial_\mu{\cal J}^\mu_{L/R} = \partial_t{\cal J}^0_{L/R}+\partial_2{\cal J}^2_{L/R} = \mp \frac{N_c}{12\pi^2}B(E\mp\epsilon) \, ,
\ee
where we have used the definition of $\mu_{5,t}$ (\ref{mut1})
and $\kappa M_{\rm KK}^2\beta\equiv N_c B/16\pi^2$.
This is exactly the consistent anomaly (\ref{consanomaly}), because
\be \label{BE12}
\left. \mp \frac{N_c}{48\pi^2}F_{\mu\nu}^{L/R}\widetilde{F}_{L/R}^{\mu\nu}\right|_{z=\infty} = \mp\frac{N_c}{12\pi^2}B(E\mp\epsilon) \, .
\ee
The new currents ${\cal J}_{L/R}^{\prime \mu}$ from the modified 
action are simply obtained
by multiplying the CS contribution of the currents (\ref{J0123sym}) by 3/2. 
Doing so in the explicit results (\ref{J0123sym}), this yields
\be
\partial_\mu{\cal J}_{L/R}^{\prime \mu} = \partial_t{\cal J}^{\prime 0}_{L/R}+\partial_2{\cal J}^{\prime 2}_{L/R} = 0 \, ,
\ee
which can also be inferred in generality from (\ref{consanomaly}).
Consequently, the anomaly has disappeared. In other words,
the new vector and the axial currents are both conserved. Nevertheless, one finds nonzero currents in the direction of the magnetic field.
Multiplying the CS contribution in eq.\ (\ref{J3sym}) by 3/2 one obtains  ${\cal J}^{\prime 3}_R+ {\cal J}^{\prime 3}_L= N_c/(4\pi^2)\,B\mu_5$  
and ${\cal J}^{\prime 3}_R- {\cal J}^{\prime 3}_L= N_c/(4\pi^2)\,B\mu$ \cite{Bergman:2008qv}, both of which are 1/2 times the
results of refs.~\cite{Fukushima:2008xe}
and \cite{Metlitski:2005pr,Newman:2005as}, respectively (cf.\ sec.\
\ref{sec:cme1} below).  

In the remainder of the paper we shall again consider the full, unmodified chiral currents (\ref{currents})
which contain the
complete covariant anomaly upon inclusion of the counterterm (\ref{DeltaS}).

\section{Axial and vector currents}
\label{sec:cme}

In this section we shall use the results from the previous sections to compute the vector and axial currents in the presence of a 
magnetic field and a quark chemical potential $\mu$ as well as an axial chemical potential $\mu_5$. 
We have seen that a consistent definition of the currents is not obvious in the given setup. We shall focus on the definition
of the currents presented in sec.\ \ref{sec:action} since they reproduce, together with Bardeen's counterterm, the correct anomaly. 
Before going into the details, let us explain the expected physics behind the vector current, in other words the chiral magnetic effect.

\subsection{The chiral magnetic effect}\label{sec:cme1}

A (noncentral) heavy-ion collision, where the chiral magnetic effect is expected to occur, is more complicated than we can capture
with our thermodynamic description. The physical situation and its simplified description within a thermodynamic approach is as follows \cite{Fukushima:2008xe}. 
In the high-temperature phase gluonic sphaleron configurations with
nonzero winding number should be produced with relatively high
probability, inducing
an imbalance in left- and right-handed quarks due to the QCD anomaly
and thus a nonzero axial number density $n_5$. 
In the simple picture applied in ref.\ \cite{Fukushima:2008xe},
such chirality changing transitions are assumed to have taken place
in a nonequilibrium situation, after which in equilibrium a finite $n_5$
is no longer changed by the QCD anomaly.
The QED anomaly on the other hand does not change $n_5$ as long as only
a magnetic field is present, so $n_5$ can be considered a conserved
quantity for which we may introduce $\mu_5$ as the corresponding chemical 
potential. (We have introduced also electric fields above for
the sake of checking the axial anomaly,
but shall set them to zero in the final results.)
Nonzero quark masses and/or nonzero chiral condensates can be expected to
lead to a decay of $n_5$. In the given context,  it is thus questionable
to apply the equilibrium description also to the chirally broken phase, and strictly speaking our approach should be extended to a 
nonequilibrium calculation.

Let us now briefly recapitulate the physics behind the
occurrence of the vector current which constitutes the chiral magnetic effect
in terms of a (quasi)particle picture \cite{Kharzeev:2007jp,Fukushima:2008xe}. 
Suppose the magnetic field leads to a spin polarization of all fermions, i.e., 
the spins of all quarks are aligned parallel or antiparallel to the magnetic field 
depending on their charge being positive or negative. Massless right-handed fermions, which have positive helicity,
have momenta parallel to their spin, so they move parallel to the magnetic field if they have positive charge, and antiparallel otherwise. 
For left-handed fermions with negative helicity, the situation is exactly reversed. 
If there are more right-handed than left-handed fermions, $n_5>0$, 
there is a resulting net electromagnetic current parallel to the magnetic field. (Antifermions have helicity opposite to chirality but 
also opposite charge,
so they give a current in the same direction.)
For weakly-coupled fermions this picture applies
since in the lowest Landau level indeed all fermions have their spins aligned in the direction of the magnetic field according to their charge. The chiral magnetic effect then
results solely from the lowest Landau level. The contribution of fermions in higher Landau levels, where both parallel and antiparallel spin
projections are populated, cancels out. This can be seen explicitly upon using the thermodynamic potential of free fermions in a magnetic field,
and the resulting current is \cite{Fukushima:2008xe}
\be \label{Jtop}
J = \frac{N_c}{2\pi^2}\mu_5 B \, .
\ee
In our model we cannot see any Landau levels directly. Therefore, let us also repeat another, apparently more general,
derivation of the chiral magnetic effect. It is based on an energy conservation argument originally 
pointed out by Nielsen and Ninomiya 
\cite{Nielsen:1983rb} and applied in ref.\ \cite{Fukushima:2008xe}. It states that an energy $2\mu_5$ is needed to replace a fermion at the left-handed Fermi surface $\mu_L$ with a fermion at the right-handed Fermi surface $\mu_R$. This conversion changes the axial number density 
by $dN_5=2$, i.e., the energy actually is $\mu_5 dN_5$. Such a change in $N_5$ is possible through the QED anomaly in the presence of 
an electric and a (non-orthogonal) magnetic field. The energy can thus be provided by an electric current. Hence, the change in $N_5$ per
unit volume and time is given by the electric power per unit volume ${\bf J}\cdot{\bf E}$, 
\be \label{NN}
{\bf J}\cdot{\bf E} = \mu_5 \frac{dn_5}{dt} \, .
\ee
Now we know from the (covariant) axial anomaly (\ref{covanomaly2}) that, 
with $\nabla\cdot {\bf J}_5=0$ and $n_5={\cal J}_5^0$, 
\be \label{dn5dt}
\frac{dn_5}{dt} = \frac{N_c}{2\pi^2}{\bf B}\cdot {\bf E} \, .
\ee
Inserting this into eq.\ (\ref{NN}) and taking ${\bf B}$ and ${\bf E}$ parallel yields a current ${\bf J}$ in the direction of ${\bf B}$,
given by eq.\ (\ref{Jtop}). Note that this argument only works for a nonzero, although arbitrarily small, electric field.

Besides the vector current we shall
also compute the axial current for which the analogous topological result is \cite{Metlitski:2005pr,Newman:2005as}
\be \label{Jtopaxial}
J_5 = \frac{N_c}{2\pi^2}\mu B \, ,
\ee
which is proportional to the ordinary quark chemical potential and thus
of potential interest in neutron and quark star physics.

\subsection{Currents with consistent anomaly}

We have already computed the currents in the symmetric phase, see eqs.\ (\ref{J0123sym}). 
The analogue for the broken phase, obtained from the definition (\ref{currents}) and the gauge fields and field strengths (\ref{gaugefields1})
and (\ref{fieldstrengths1}) is
\begin{subequations} \label{J0123broken}
\bea
T<T_c: \quad {\cal J}^0_{L/R} &=& \pm 4\kappa M_{\rm KK}^2\beta\left[-\mu_{5,t}\coth\beta\pi\mp(\jmath_t-\epsilon t)
+\frac{Et\pm (\jmath_t-\epsilon t)}{3}\right] \, , \label{J0broken}\\
{\cal J}^1_{L/R} &=& 0 \, , \\
{\cal J}^2_{L/R} &=& \pm \frac{4\kappa M_{\rm KK}^2}{3} \beta \,x_2 \left[E\mp\epsilon\frac{\beta\coth\beta\pi}{2\rho(\beta)} \right] \, , \label{J2broken} \\
{\cal J}^3_{L/R} &=& \mp 4\kappa M_{\rm KK}^2\beta\left[\mp \mu_{5,t}-(\jmath_t-\epsilon t)\coth\beta\pi-\frac{\mu_t\mp\mu_{5,t}}{3}\right] \, . 
\label{J3broken} 
\eea
\end{subequations}
Again, to make the origin of the various terms transparent we have written the CS contributions separately. All terms containing a 1/3 come from 
the CS action. 
As for the symmetric phase, we can easily check the consistent anomaly (\ref{consanomaly}). Using the expression for the supercurrent
(\ref{jmin}) and $\kappa M_{\rm KK}^2\beta\equiv N_c B/16\pi^2$ we find
\be
\partial_t {\cal J}^0_{L/R} +\partial_2{\cal J}^2_{L/R} = \mp \frac{N_c}{12\pi^2}B\left(E\mp \epsilon\frac{\beta\coth\beta\pi}{2\rho(\beta)}\right)
\ee
and
\be
\mp \frac{N_c}{48\pi^2}F_{\mu\nu}^{L/R}\widetilde{F}_{L/R}^{\mu\nu} = 
\mp \frac{N_c}{12\pi^2}B\left(E\mp \epsilon\frac{\beta\coth\beta\pi}{2\rho(\beta)}\right) \, ,
\ee
which confirms eq.\ (\ref{consanomaly}). The axial electric field seems to be modified by a complicated function of the dimensionless
magnetic field. This originates from the mixing of the electric field with the supercurrent, which both enter the boundary value of 
$A_3$. We shall see that this somewhat strange structure disappears after adding Bardeen's counterterm.

\FIGURE[t]{\label{figratio}
{ \centerline{\def\epsfsize#1#2{0.7#1}
\epsfbox{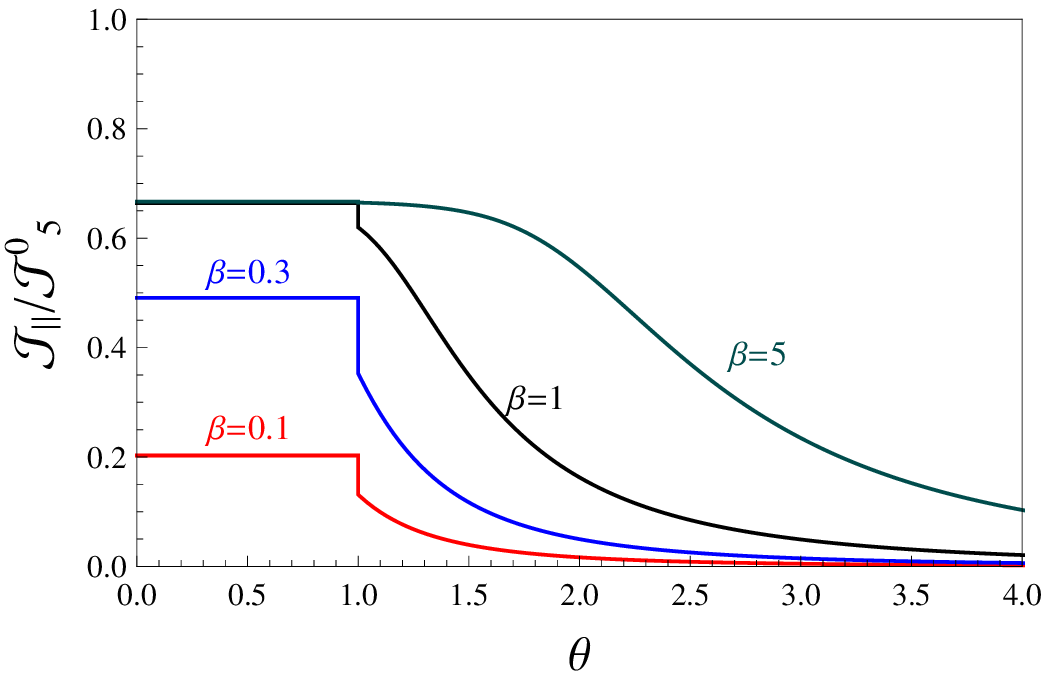}
\epsfbox{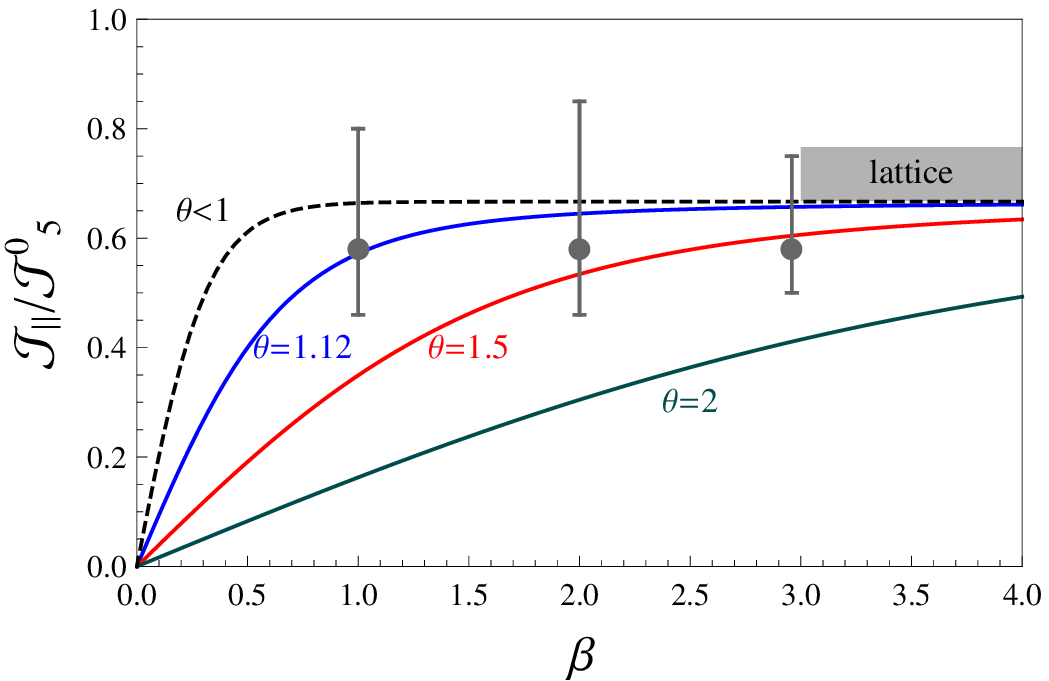}}
}
\caption{Vector current $\mathcal J_\parallel={\cal J}_3$ 
per imbalance of right- and left-handed fermions $n_5={\cal J}_5^0$
as a function of the dimensionless temperature $\theta=2\pi T/M_{\rm KK}$ for different values
of the dimensionless magnetic field $\beta = \alpha B/M_{\rm KK}^2 \simeq B/(0.35\,{\rm GeV}^2) \simeq B/(2\cdot 10^{19}\,{\rm G})$ 
(left panel) and as a function of $\beta$ for different values 
of $\theta$ (right panel). The critical temperature for the chiral phase transition is $T_c=M_{\rm KK}/(2\pi)$, i.e., $\theta_c =1$.
The currents in this plot are obtained using the consistent anomaly, i.e., {\it before} adding Bardeen's counterterm
to fulfill the covariant anomaly. After this term is added, the vector current vanishes exactly. 
The left plot shows the discontinuity at the first order chiral phase transition. This discontinuity vanishes for asymptotically large magnetic 
fields. The right panel shows that the current saturates at a value of ${\cal J}_\parallel=\frac23 {\cal J}_5^0$, in very good agreement with the 
lattice data for the root mean square value of fluctuations of vector currents and axial densities
\cite{Buividovich:2009wi}. The three lattice data points are taken from from figs. 4 and 8 of 
ref.\ \cite{Buividovich:2009wi} and correspond to a temperature $T=1.12\,T_c$. 
The shaded area indicates the results read off from fig.\ 11 of ref.\ \cite{Buividovich:2009wi} for the cleaner case of a ($T=0$) 
instanton-like configuration, where
the corresponding points lie between ${\cal J}_\parallel/{\cal J}_5^0 \simeq 0.66 - 0.77$ for magnetic fields of 
$\beta \simeq 3.0$ and higher.}
}

From the results (\ref{J0123sym}) and (\ref{J0123broken}) we may compute the 
vector currents in the chirally symmetric and broken phases. For the following results we set $E=\epsilon=0$.
We find the same result for both phases which is
\be
{\cal J}_3 = 
(\mathcal J_{\rm YM}+
\mathcal J_{\rm CS})_3=\left(1-\frac13\right)\frac{N_c}{2\pi^2}B\mu_5 \, .
\ee
This differs by a factor 2/3 from the topological result (\ref{Jtop}). This difference is not surprising since we have not 
implemented the covariant anomaly yet. To this end we must add Bardeen's counterterm. Before doing so we 
point out an interesting result which we obtain by considering the ratio of the vector current over the axial density. From 
eqs.\ (\ref{J0123sym}) and (\ref{J0123broken}) we obtain 
\be
\frac{{\cal J}_3}{{\cal J}_5^0} = \frac{2}{3} 
\left\{\begin{array}{cc} \displaystyle{\frac{2\beta}{\theta^3}\frac{q_0}{p_0}} & \;\;\;\;{\rm for}\; T>T_c \\[2ex]
\tanh\beta\pi  &\;\;\;\; {\rm for} \; T<T_c   
\end{array}\right. \, ,
\ee
which is displayed in fig.\ \ref{figratio}. In the left panel we see that the first order chiral phase transition manifests
itself in the discontinuity of the ratio ${\cal J}_3/{\cal J}_5^0$. Interestingly, the jump vanishes for asymptotically large magnetic fields.
The curves for the symmetric phase are in qualitative agreement with the weak-coupling results in fig.\ 2 of ref.\ \cite{Fukushima:2008xe}. 
The right panel shows an intriguing agreement of our result
for the ratio ${\cal J}_3/{\cal J}_5^0$
with recent lattice results \cite{Buividovich:2009wi}
for the root mean square values of electric currents and chiral densities
at large magnetic fields.
While the very good numerical agreement might be a coincidence, the lattice results as well as our result clearly show an
asymptotic value significantly smaller than 1. If it were 1, the entire imbalance ${\cal J}_5^0$ in right-and left-handed fermions, i.e., 
all excess right-handed fermions, would contribute to the current for asymptotically large magnetic fields. This is expected at least at 
weak coupling. In this case, for sufficiently large magnetic fields, all fermions populate the lowest Landau level. Consequently, 
since the current originates solely from the lowest Landau level, as explained above, one expects
${\cal J}_3/{\cal J}_5^0 \to 1$. This is confirmed in the weak-coupling calculation of ref.\ \cite{Fukushima:2008xe}, see fig.\ 6 in this reference. 
The lattice result suggests that at strong coupling there may be important
modifications to the Landau level picture.
We emphasize, however, that fig.\ \ref{figratio} is not yet our final physical prediction. The model has not yet been 
appropriately renormalized in order to exhibit the covariant anomaly. 

We also remark that the scale of our magnetic fields 
is very large such that for all physical applications, be it in heavy-ion collisions or in magnetars, the limit of weak magnetic fields is 
sufficient. In fact, a dimensionless magnetic field $\beta = 1$ corresponds roughly to a magnetic field
$B \simeq 2 \cdot 10^{19}\, {\rm G}$ if one follows refs.\ \cite{Sakai:2004cn,Sakai:2005yt} and sets $N_c=3$, 
$M_{\rm KK}\simeq 949\, {\rm MeV}$, $\kappa \simeq 0.00745$, which
fits the experimental values for the pion decay constant and the rho meson mass. Therefore,
for all applications we have in mind, $\beta\ll 1$. Moreover, one should recall that we have used the YM approximation for the DBI action. This 
is of course a good approximation for small magnetic fields, but our extrapolation to larger magnetic fields may be subject to modification 
when the full DBI action is employed. On the other hand, in the limit $\beta\gg 1$, the results for the on-shell action, 
eqs.\ (\ref{YMCSonshell}) and (\ref{YMCSOmega2}) exhibit a strong suppression of the YM action compared to the CS action. This suggests that 
our approximation is reliable also for asymptotically large magnetic fields.

\subsection{Currents with covariant anomaly and absence of the chiral magnetic effect}

The next step is to include
Bardeen's counterterm (\ref{DeltaS}) in order to implement the covariant anomaly. In the broken phase there is a slight complication because
the counterterm should only involve genuine
background gauge fields, and not those boundary values of the
bulk gauge fields that due to the gauge choice $A_z=0$
represent gradients of the pion field.
This means that we have to subtract the time-independent part of the supercurrent $\jmath=-\beta\mu/2\rho$ from the boundary 
values of the $A_3^{L/R}$. Then, with eq.\ (\ref{DeltaJ}) and the value of $c$ from eq.\ (\ref{c}), the contributions of the counterterm to the currents are
\begin{subequations} \label{DeltaJ0123broken}
\bea
T<T_c: \quad \Delta{\cal J}^0_{L/R} &=& \pm \frac{2 \kappa M_{\rm KK}^2}{3}\beta [3(A_3^{R/L}\mp\jmath)-(A_3^{L/R}\pm\jmath)] \non
& =& \mp\frac{4\kappa M_{\rm KK}^2}{3}\beta t\left[ E \pm 2\epsilon\frac{\beta\coth\beta\pi}{2\rho(\beta)}\right]  \, , \label{DeltaJ0broken}\\
\Delta{\cal J}^1_{L/R} &=& 0 \, , \\ 
\Delta{\cal J}^2_{L/R} &=& \mp \frac{4\kappa M_{\rm KK}^2}{3}\beta \,x_2\left[E\mp\frac{\beta\coth\beta\pi}{2\rho(\beta)}\right]  \, , \label{DeltaJ2broken} \\
\Delta{\cal J}^3_{L/R} &=& \mp \frac{2\kappa M_{\rm KK}^2}{3}\beta(3A_0^{R/L}-A_0^{L/R}) 
= \mp \frac{4\kappa M_{\rm KK}^2}{3}\beta(\mu_t\pm 2\mu_{5,t}) \, . \label{DeltaJ3broken}
\eea
\end{subequations}
The first observation is that the 2-component of the current vanishes after adding the counterterm. As mentioned above,
this 2-component was unphysical anyway. 
\TABLE[t]{
\begin{tabular}{|c||c|c|c|c|} 
\hline
\rule[-1.5ex]{0em}{5ex} & $\bar{\cal J}^0$ & $\bar{\cal J}_5^0$ & $\;\;$$\bar{\cal J}_\parallel$$\;\;$ & 
$\bar{\cal J}_\parallel^5$  \\ \hline\hline
\rule[-1.5ex]{0em}{6ex} $\;\;$ $T>T_c$ $\;\;$  & $\;\;$$\displaystyle{\frac{N_c}{4\pi^2}\mu B\,\frac{\theta^3}{\beta}\frac{p_0}{q_0}}$$\;\;$ & 
$\;\;$$\displaystyle{\frac{N_c}{4\pi^2}\mu_5 B\,\frac{\theta^3}{\beta}\frac{p_0}{q_0}}$$\;\;$  & 0 
& $\;\;$$\displaystyle{\frac{N_c}{2\pi^2}\mu B}$$\;\;$  \\[2ex] \hline
\rule[-1.5ex]{0em}{6ex} $\;\;$ $T<T_c$ $\;\;$ & $\;\;$ $\displaystyle{\frac{N_c}{6\pi^2}\mu B\,\frac{\beta}{\rho}}$$\;\;$ &
$\;\;$ $\displaystyle{\frac{N_c}{2\pi^2}\mu_5 B\,\coth\beta\pi}$$\;\;$ &   0 & $\;\;$$\displaystyle{\frac{N_c}{4\pi^2}\mu B\,
\frac{\beta\coth\beta\pi}{\rho}}$$\;\;$ 
\\[2ex] \hline
\end{tabular}
\caption{Vector and axial densities $\bar{\cal J}^0$, $\bar{\cal J}_5^0$, and vector and axial currents 
$\bar{\cal J}_\parallel$, $\bar{\cal J}_\parallel^5$ in the direction of the magnetic field $B$ after adding Bardeen's counterterm.
All results are given as functions of the dimensionless
temperature $\theta=2\pi T/M_{\rm KK}$ and the dimensionless magnetic field $\beta= \alpha B/M_{\rm KK}^2$. The densities in the 
chirally symmetric phase ($T>T_c$) depend on temperature; the ratio $p_0/q_0$ behaves as $p_0/q_0\to 1$ for $\beta/\theta^3\to 0$ and 
$p_0/q_0\to 2\beta/\theta^3$ for $\beta/\theta^3\to\infty$. In the chirally broken 
phase ($T<T_c$), all quantities are independent of temperature; the function $\rho$  behaves as $\rho\to 3/(2\pi)$ for $\beta\to 0$ and 
$\rho\to \beta$ for $\beta\to \infty$. 
The vector current vanishes exactly in both symmetric and broken phases; this indicates 
the absence of a chiral magnetic effect in the Sakai-Sugimoto model, see discussion in the text. 
For the axial current, the temperature-independent topological 
result is reproduced in the symmetric phase. See fig.\ \ref{figaxial} for the comparison of the axial currents in the symmetric and broken phases.
 }
\label{tablecurrents}
}
The cancellation of this component is therefore, besides the covariant anomaly, another sign for the 
necessity of the counterterm. The covariant anomaly is now correctly
contained in the renormalized 
currents $\bar{\cal J}^\mu_{L/R}={\cal J}^\mu_{L/R}+
\Delta{\cal J}^\mu_{L/R}$. This is clear by construction, and can also
be verified explicitly: adding eqs.\ (\ref{DeltaJ0123broken}) to eqs.\ (\ref{J0123broken}),
yields 
\be \label{anomalyexplicit}
\partial_\mu \bar{\cal J}^\mu = 0 \, , \qquad \partial_\mu \bar{\cal J}^\mu_5 = \partial_t \bar{\cal J}^0_5 = \frac{N_c}{2\pi^2}BE
\, ,
\ee
with the vector and axial currents $\bar{\cal J}^\mu$, $\bar{\cal J}^\mu_5$. 

The contributions of the counterterm to the currents in the symmetric phase are 
\begin{subequations} \label{DeltaJ0123sym}
\bea
T>T_c: \quad \Delta{\cal J}^0_{L/R} &=& \pm \frac{2 \kappa M_{\rm KK}^2}{3}\beta (3A_3^{R/L}-A_3^{L/R}) =
 \mp\frac{4\kappa M_{\rm KK}^2}{3}\beta t (E \pm 2\epsilon)   \, , \label{DeltaJ0sym}\\
\Delta{\cal J}^1_{L/R} &=& 0 \, , \\ 
\Delta{\cal J}^2_{L/R} &=& \mp \frac{4\kappa M_{\rm KK}^2}{3}\beta \,x_2(E\mp\epsilon)  \, , 
\label{DeltaJ2sym} \\
\Delta{\cal J}^3_{L/R} &=& \mp \frac{2\kappa M_{\rm KK}^2}{3}\beta(3A_0^{R/L}-A_0^{L/R}) 
= \mp \frac{4\kappa M_{\rm KK}^2}{3}\beta(\mu_t\pm 2\mu_{5,t}) \, . \label{DeltaJ3sym}
\eea
\end{subequations}
These counterterms have to be added to the currents (\ref{J0123sym}). Again, the 2-component of the currents is canceled, and the covariant anomaly
can again be verified explicitly. 

\FIGURE[t]{\label{figaxial}
{ \centerline{\def\epsfsize#1#2{0.8#1}
\epsfbox{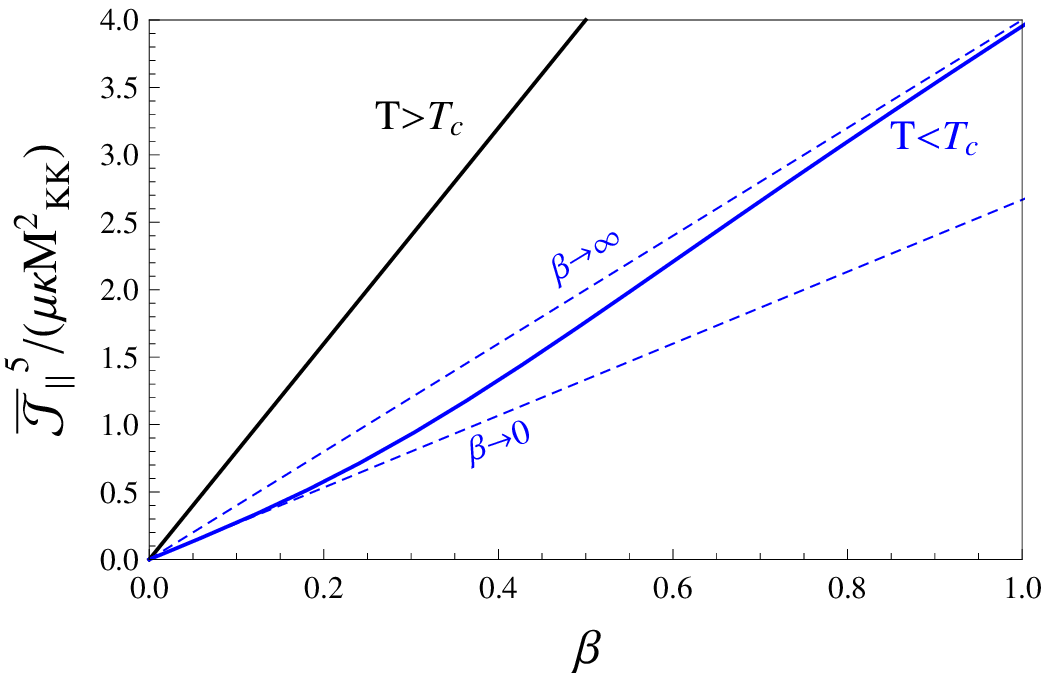}}
}
\caption{Axial current $\bar{\cal J}^5_\parallel$ in chirally symmetric ($T>T_c$) and chirally broken ($T<T_c$) phases in the presence of 
chemical potential $\mu$ as functions of the dimensionless magnetic field 
$\beta= \alpha B/M_{\rm KK}^2$, i.e., the magnetic field in units of $M_{\rm KK}^2/\alpha \simeq 2\cdot 10^{19}\,{\rm G}$. 
In the symmetric phase the current is linear in $B$, while the current
in the broken phase is linear only for small $B$ and asymptotically large $B$ as indicated by the dashed lines. Due to the huge
scale for the magnetic field, the small-$B$ approximation for the axial current is sufficient for astrophysical applications. In this case 
the current in the broken phase is simply 1/3 of the current in the symmetric phase.   
The analytical results are given in table \ref{tablecurrents} where it is also shown that the vector current vanishes.
 }  
}

The results for the currents after setting $E=\epsilon=0$ are given in table \ref{tablecurrents} for both the
symmetric and the chirally broken phase.
For the axial current we find that the counterterm exactly cancels
the CS part of the current,
\be
\bar{\mathcal J}_5^3={\mathcal J}_5^3+\Delta {\mathcal J}_5^3=
({\mathcal J}_{\rm YM})_5^3\,.
\ee
In the chirally symmetric phase, this yields
exactly the expected topological result (\ref{Jtopaxial}).
In the broken phase, the current is suppressed (but nonvanishing, in
contrast to the results obtained with a modified CS action \cite{Bergman:2008qv}). To lowest order in the magnetic fields as well as for asymptotically large magnetic fields this suppression is simply given by 
a numerical factor. For intermediate magnetic fields the difference to the symmetric phase is given by a complicated function of $B$. 
We plot this result in fig.\ \ref{figaxial}. 

The most striking of our results shown in table \ref{tablecurrents}
is that 
for both phases the renormalized vector current is zero for all magnetic fields, 
\be\label{J30}
\bar{\mathcal J}_3=
({\mathcal J}_{\rm YM}+{\mathcal J}_{\rm CS}+
\Delta {\mathcal J})_3=\left(1-\frac13-\frac23\right)\frac{N_c}{2\pi^2}B\mu_5
=0,
\ee
i.e., the chiral magnetic effect 
has completely disappeared after adding Bardeen's counterterm. 
The vector current has been computed in the Sakai-Sugimoto model before, and both existing (but differing) results are nonvanishing.
One of the results \cite{Bergman:2008qv} is 1/2 of the topological result (\ref{Jtop}).
This result, however, has been obtained with the modified action discussed in sec.\ \ref{sec:currents} which amounts to multiplying the CS contribution
by 3/2 (and leaving out the counterterm). 
As we have seen, this modified action leads
to a vanishing anomaly. 
Another result has recently been presented in ref.\ \cite{Yee:2009vw} as a limit case of a more general frequency-dependent calculation, but using
only the YM part of the current. This gives the topological result (\ref{Jtop}), as can also be seen from eq.\ (\ref{J30}). 
However, as we have shown, this does not produce the complete covariant anomaly, see eqs.\ (\ref{YMcurranomalies}).

\newcommand\V{F_V\widetilde F_V }
\newcommand\A{F_A\widetilde F_A }
\newcommand\VA{F_V\widetilde F_A }


\renewcommand{\arraystretch}{1.2}
\TABLE[ht]{
\begin{tabular}{|c||c|c|c|c|} 
\hline
~ & $\mathcal J_{\rm YM}$ & $\mathcal J_{\rm YM+CS}$ & $\mathcal J_{\rm YM+CS}+\Delta \mathcal J$ & $\mathcal J'_{\rm YM+CS}$ \\
\hline\hline
anomaly & ``semi-covariant'': & consistent: & \underline{covariant:} & absent: \\
\hline
\rule[-1.5ex]{0em}{4.5ex} 
$\partial_\mu \mathcal J^\mu_5 / \frac{N_c}{24\pi^2}$ & 
 \underline{$3 \V$}$+3\A$ & $\V+\A$ & \underline{$3\V+\A$} & 0 \\[0.5ex]
\hline
\rule[-1.5ex]{0em}{4.5ex} 
$\partial_\mu \mathcal J^\mu / \frac{N_c}{24\pi^2}$ & 
 $6\VA$ & $2\VA$ & \underline{$0$} & $0$ \\[0.5ex]
\hline\hline
\rule[-1.5ex]{0em}{4.5ex} $\left.(\mathcal J_\parallel^5 / \frac{\mu B N_c}{2\pi^2})\right|_{T>T_c}$ & 
 1 & $\frac23$ & 1 & $\frac12$ \\[1ex]
\hline
\rule[-1.5ex]{0em}{4.5ex} 
$\mathcal J_\parallel / \frac{\mu_5 B N_c}{2\pi^2}$ & 
 1 & $\frac23$ & 0 & $\frac12$ \\[0.5ex]
\hline
\rule[-1.5ex]{0em}{4.5ex} 
$\left.(\mathcal J_\parallel/\mathcal J^0_5)\right|_{B\to\infty}$  & 
 1 & $\frac23$ & 0 & $\frac12$ \\
\hline
\end{tabular}
\caption{Summary of results for the different (parts of the) chiral currents:
the Yang-Mills part $\mathcal J_{\rm YM}$ (exclusively considered
in ref.~\cite{Yee:2009vw}), the complete current prior to
renormalization $\mathcal J_{\rm YM+CS}$ ($\equiv \mathcal J$ in the text), 
the complete current plus
Bardeen's counterterm $\mathcal J_{\rm YM+CS}+\Delta \mathcal J\equiv\bar{\mathcal J}$, and
the chiral current obtained by modifying the Chern-Simons action
according to ref.~\cite{Bergman:2008qv,Lifschytz:2009si}, $\mathcal J'_{\rm YM+CS}$.
The correct result for the covariant anomaly is underlined.
A ``1'' in the results for the axial current $\mathcal J_\parallel^5$ means agreement with the
exact QCD result of ref.~\cite{Metlitski:2005pr,Newman:2005as}; 
a ``1'' in the results
for the electromagnetic current $\mathcal J_\parallel$ means agreement with the weak-coupling approach of ref.~\cite{Fukushima:2008xe}. 
}\label{tab:summary}
}

One of the purposes of our paper is to point out the differences of these results and the problems of the various approaches 
regarding the correct anomaly. A summary of our findings is given in
table \ref{tab:summary}.
Although in our approach the correct anomaly is ensured, we do not claim to have the final answer since the problem of the ambiguity 
of the currents, see sec.\ \ref{sec:currents}, remains. 
Our approach shows that the CS part of the currents is important for two reasons. First, as realized already in earlier works 
\cite{Rebhan:2008ur,Hata:2008xc,Hashimoto:2008zw}, it naturally gives a nonzero contribution when the 
currents are defined by varying the full action. 
Second, and maybe more importantly, only by including the CS contribution does one reproduce the standard result for the consistent anomaly. 
And only then one can completely implement the covariant anomaly 
(i.e., a conserved vector current even in the presence of axial gauge fields)
by adding an appropriate counterterm as a holographic renormalization. 
We have explained why this
counterterm, even in the absence of axial field strengths, but in the presence of an axial chemical potential, 
changes the result for the vector current. We do not, however, see a general 
reason why the counterterm must render the vector current zero, i.e., why by requiring the current to be conserved the current itself should
disappear as it turned out to be the case in our explicit calculation.

After having understood the difference of our result to previous results in the same model, let us discuss its significance in view of the 
apparent contradiction to the result (\ref{Jtop}). As explained above, this result can be derived by using the 
Landau-level structure of fermions in a magnetic field. One might thus view our result as an indication that there are 
no fermionic quasiparticles filling 
Landau levels in the Sakai-Sugimoto model. This may be particularly interesting in view of the recent attempts to see Landau-level-like
structures in holographic models \cite{Basu:2009qz,Denef:2009yy}. However, as we have pointed out, the derivation of the chiral magnetic effect
via the energy conservation argument by Nielsen and Ninomiya appears to be more general. 
Obviously, the energy conservation (\ref{NN}) does not hold with our results because the 
left-hand side is zero while the right-hand side yields the expected nonzero result from the anomaly, see eq.\ (\ref{anomalyexplicit}).
More precisely, one can check that eq.\ (\ref{NN}) holds before adding Bardeen's counterterm while the counterterm itself violates
eq.\ (\ref{NN}). However, the form of the counterterm
seems to be uniquely determined by the requirements of
parity and the possibility of accommodating it in holographic renormalization.
This raises the question whether the apparently general energy argument actually uses properties of the system which are different in our
strong-coupling approach. Clearly, also in our system, chirality is converted by a rate simply given 
by the anomaly. Possibly the energy needed for this conversion cannot be written as in eq.\ (\ref{NN}). A reason might be that this
energy makes use of the existence of Fermi surfaces for the right- and left-handed particles which are absent in our model. 
It is tempting to speculate that the chiral magnetic effect indeed vanishes in the strong-coupling limit and that the weak-coupling
results together with the recent observations of charge separation in heavy-ion collisions suggest that the quark-gluon plasma generated in such a 
collision is sufficiently weakly coupled to exhibit the chiral magnetic effect. A deeper understanding of our result, however, 
seems required before drawing this conclusion.

We recall that in the context of heavy-ion collisions the magnetic field clearly is time-dependent, in contrast to our assumption 
of a constant magnetic field. Therefore, in order to compute the induced current, one has to consider the frequency-dependent 
chiral conductivity \cite{Kharzeev:2009pj,Yee:2009vw}, whereas
our result only corresponds to the zero-frequency limit\footnote{
Judging from the calculation of the chiral magnetic conductivity in 
ref.\ \cite{Yee:2009vw} (where only the YM part was taken into account), one might expect that the full result, taking into 
account also the (frequency-independent) 
CS part and Bardeen's counterterm, leads to a nonzero conductivity for asymptotically large frequency \cite{priv}. 
This seems curious, although we do not see a fundamental 
reason for this to be unphysical.}. In other words, even if the conductivity at zero frequency vanishes, 
a nonzero (time-dependent) current can be expected if there is a nonvanishing conductivity at nonzero frequencies. However, this does not imply  
electric charge separation because the separation of charges is proportional to the zero-frequency limit of the conductivity
\cite{Kharzeev:2009pj}. This is easy to understand in analogy to a capacitor which cannot be charged with an alternating current, i.e., 
integrating the induced current over time will lead to a vanishing charge separation as long as there is no direct current.

\section{Summary and conclusions}
\label{sec:conclusions}

We have studied the strong-coupling behavior of chiral fermions in the presence of a chemical potential and a background 
magnetic field in the chirally broken and the chirally symmetric phases. To this end, we have used the Sakai-Sugimoto model which is the model 
that at present comes closest to providing a gravity dual to (large-$N_c$) QCD. Our focus was the calculation of 
(topological) axial and vector currents, which are direct consequences of the anomalies in the model. 

In particular, we have investigated the chiral magnetic effect, which has been studied previously 
in a weak-coupling approach \cite{Fukushima:2008xe} and on the lattice \cite{Buividovich:2009wi}. We have 
pointed out that a reliable calculation within the Sakai-Sugimoto model requires a careful discussion of the QED anomaly in the model. 
The standard value of the {\it consistent} anomaly arises naturally in the model for the most straightforward definition of the current. 
For this result it is crucial to include the contributions from the CS term which are sometimes ignored in the literature.
The {\it covariant} anomaly can then be implemented by adding Bardeen's counterterm \cite{Bardeen:1969md}, which is also known to be required
in chiral models with a Wess-Zumino-Witten term \cite{Kaymakcalan:1983qq}, and we have pointed out that this (finite) counterterm has a form that
appears consistent with the procedure of holographic renormalization.

After these general discussions we have solved the equations of motion for the chirally broken and the chirally symmetric phases explicitly.
In our approximation of the DBI action to lowest order in the gauge fields, the solutions are completely
analytical. Electric (vector and axial) fields parallel to the magnetic field have been considered in order to check the anomaly explicitly,
but they are not needed and set to zero
for our physical (equilibrium) results, which only require
magnetic fields in the presence of chemical potentials.

In the presence of a quark chemical potential and a large magnetic field, we have calculated the axial current,
which may be of interest for astrophysical phenomena such as pulsar kicks
\cite{Gorbar:2009bm}. In the chirally symmetric
phase we have reproduced the known topological result \cite{Metlitski:2005pr,Newman:2005as}, while in the chirally broken phase, 
the current has turned out to be smaller but nonvanishing. These results can also be obtained by using only the YM part of the current, i.e., in the case of the axial current the CS contribution
and the contribution of Bardeen's counterterm cancel each other.

This is different for the vector current. In this case, only the YM contribution yields a result 
in agreement with ref.\ \cite{Fukushima:2008xe}. With the full current, and after adding Bardeen's counterterm, the vector current 
becomes zero, indicating the absence of the chiral magnetic effect. 
This is in no obvious contradiction to the result obtained 
in the weak-coupling approach
\cite{Fukushima:2008xe}, since at weak coupling the chiral magnetic effect is a phenomenon that originates 
solely from fermionic quasiparticles in the lowest Landau level, and it is not clear whether this structure
persists at strong coupling.

In comparison to the result
from recent lattice calculations \cite{Buividovich:2009wi} we have pointed out an intriguing agreement {\it before} 
adding Bardeen's counterterm, i.e., within the 
consistent anomaly. There the
vector current per chirality approaches approximately the value 2/3 for asymptotically 
large magnetic field. This is clearly different from the weak-coupling approach where this ratio approaches 1. This raises the 
question whether this asymptotic value can distinguish between strong and weak coupling. It also raises the question whether the lattice
result relates to the consistent, as opposed to the covariant, anomaly. 

Although we have implemented the correct covariant anomaly, 
also in our approach 
a problem remains. Namely, upon computing the free energy explicitly and then taking the derivative with respect to the appropriate 
source, the currents turn out to be different from the straightforward definition via the gauge/gravity correspondence. 
This somewhat disturbing discrepancy can be attributed to boundary terms at spatial infinity. 
We have discussed that a previously suggested fix of this problem by modifying the action \cite{Bergman:2008qv} 
seems to be not acceptable because it entirely eliminates the axial anomaly
from the correspondingly modified currents. Only the YM part of these currents
are anomalous, but those suffer from the same thermodynamic inconsistency that this modification was meant to fix.
Because of these ambiguities, further studies are clearly needed
(see also ref.\ \cite{Hata:2008xc} for other issues concerning the definition
of chiral currents in the Sakai-Sugimoto model). 

Quantitative improvements could be achieved by extending 
our calculation to the full DBI action, though they should be minor for magnetic field strengths
of practical interest. More critical, but also considerably more difficult,
would be the generalization of our ansatz to allow for inhomogeneous field
configurations and/or inhomogeneous solutions. This might be required
to resolve the ambiguities in the definition of the chiral currents that we
have discussed, since those are related to spatial surface contributions
in the CS action.
With our present definition of the chiral currents we have been led to
question the very 
existence of the chiral magnetic effect in the strong-coupling
regime of the Sakai-Sugimoto model (which is gravity dual 
to large-$N_c$ QCD only in its inaccessible weak coupling limit). 
In this context it would be important to understand whether
in the strong-coupling regime one has
Landau-level-like structures, 
as conjectured in ref.\ \cite{Thompson:2008qw}. 
This is interesting also in view of recent studies in different gauge/gravity models \cite{Basu:2009qz,Denef:2009yy}.

\acknowledgments

We thank especially K.\ Landsteiner, I.\ Shovkovy and H.\ Warringa for very valuable discussions. 
We also acknowledge helpful comments from M.\ Chernodub, D.\ Kharzeev, G.\ Lifschytz, E.\ Lopez, M.\ Kaminski, D.\ Son, and H.-U.\ Yee.
This work has been supported by FWF project no.\ P19958.

\appendix 

\section{Solving the equations of motion}
\label{app:solving}

\subsection{Chirally broken phase}
\label{app:broken}

In this appendix we solve the equations of motion in the broken phase, eqs.\ (\ref{E123}).
The equation of motion for $A_z$ (\ref{E3}) is trivially integrated with respect to time $t$ to yield 
\be \label{dzA0}
k\partial_z A_0 = -2\beta te(z) +k\partial_z \widetilde{A}_0 \, , 
\ee
where we have denoted $e(z)\equiv -\partial_t A_3$ and where 
we have written the $t$-independent integration constant as $k\partial_z \widetilde{A}_0$, to be determined below. 
Inserting this into eqs.\ (\ref{E1}) and (\ref{E2})
yields
\begin{subequations}
\bea
\partial_z(k\partial_z\widetilde{A}_0)&=&2\beta\partial_z A_3 +2\beta t\partial_z e \, , \label{E11}\\
\partial_z(k\partial_z A_3)&=&2\beta\partial_z \widetilde{A}_0 -(2\beta)^2t\frac{e(z)}{k(z)} \, . \label{E21}
\eea
\end{subequations}
Since the left-hand side of eq.\ (\ref{E11}) does not depend on $t$, the right-hand side must be independent of $t$ too which 
implies
\be \label{dzA3}
\partial_z A_3 = -t\partial_z e +\partial_z\widetilde{A}_3 \, , 
\ee
where we have written the $t$-independent part as $\partial_z \widetilde{A}_3$. Consequently, eqs.\ (\ref{E11}) and (\ref{E21}) become
\begin{subequations} \label{eom2}
\bea
\partial_z(k\partial_z\widetilde{A}_0)&=&2\beta\partial_z \widetilde{A}_3 \, ,\\
\partial_z(k\partial_z \widetilde{A}_3)&=&2\beta\partial_z \widetilde{A}_0 -t\left[(2\beta)^2\frac{e(z)}{k(z)}-\partial_z(k\partial_z e)\right] \, .
\label{E22}
\eea
\end{subequations}
Now the square bracket on the right-hand side of eq.\ (\ref{E22}) must vanish because all other terms in the equation do not depend on $t$. 
This yields a differential equation for $e(z)$. Since $e(z)=-\partial_t A_3$, the boundary conditions for $A_3$ (\ref{A3inf}) imply 
$e(\pm\infty)=E\mp(\epsilon -\jmath_1)$, where we have decomposed the supercurrent as
\be \label{jmatht}
\jmath_t = \jmath + \jmath_1 t \, ,
\ee
with $\jmath$, $\jmath_1$ being $t$-independent. 
With these boundary conditions the equation for $e(z)$ is solved by 
\be
e(z) = E\frac{\cosh(2\beta\arctan z)}{\cosh\beta\pi}-(\epsilon-\jmath_1)\frac{\sinh(2\beta\arctan z)}{\sinh\beta\pi} \, .
\ee
To find the solution for $A_0$ and $A_3$ we first conclude from eqs.\ (\ref{dzA0}) and (\ref{dzA3}),
\begin{subequations} \label{Atz}
\bea
A_0(t,z) &=& \widetilde{A}_0(z)+ g_0(t) \non
&&-t\left[E\frac{\sinh(2\beta\arctan z)}{\cosh\beta\pi}-(\epsilon-\jmath_1)\frac{\cosh(2\beta\arctan z)}{\sinh\beta\pi}\right] \, , \\
A_3(t,z) &=& \widetilde{A}_3(z)
-t\left[E\frac{\cosh(2\beta\arctan z)}{\cosh\beta\pi}-(\epsilon-\jmath_1)\frac{\sinh(2\beta\arctan z)}{\sinh\beta\pi}\right]  \, . 
\eea
\end{subequations}
From the $z$-integration in eq.\ (\ref{dzA0}) we have obtained a $t$-dependent integration constant $g_0(t)$. Such a constant is not permissible
in $A_3$ because of the constraint $e(z)=-\partial_t A_3$. Integration constants independent 
of $z$ {\em and} $t$ are included in $\widetilde{A}_0(z)$, $\widetilde{A}_3(z)$. We shall fix $g_0(t)=\jmath_1 t\coth\beta\pi$. This removes the 
supercurrent from the vector boundary value of $A_0(t,z)$. We cannot at the same time remove the axial field $\epsilon$ from this boundary value.
This becomes clear in hindsight after determining $\jmath_t$ from minimization of the free energy. Only with the given choice of $g_0(t)$
this minimization leads to a consistent, i.e., time-independent, result for $\jmath$, $\jmath_1$. 
It is thus unavoidable for the boundary values of $A_0(t,z)$ to become time-dependent,  
\be \label{A0inf}
A_0(t,z=\pm\infty) = \mu_t \mp \mu_{5,t} \, , 
\ee
where we defined
\be 
\mu_{t}\equiv \mu+t\epsilon\coth\beta\pi \, , \qquad \mu_{5,t}\equiv \mu_5+tE\tanh\beta\pi \, .
\ee
We have now reduced the equations of motion (\ref{eom2}) to equations for $\widetilde{A}_0(z)$, $\widetilde{A}_3(z)$ 
which are simply the gauge fields in the absence of an electric field. These equations 
can be solved in general, 
\begin{subequations}
\bea
\widetilde{A}_0(z) &=& a_0 -\frac{c}{2\beta}e^{-2\beta\arctan z} + \frac{d}{2\beta}e^{2\beta\arctan z} \, , \\
\widetilde{A}_3(z) &=& a_3 +\frac{c}{2\beta}e^{-2\beta\arctan z} + \frac{d}{2\beta}e^{2\beta\arctan z} \, , 
\eea
\end{subequations}
with integration constants $a_0$, $a_3$, $c$, and $d$ which are fixed by the boundary conditions 
$\widetilde{A}_0(\pm\infty) = \mu \mp \mu_5$, $\widetilde{A}_3(\pm\infty) = \mp \jmath$. 
The resulting gauge fields $\widetilde{A}_0(z)$, $\widetilde{A}_3(z)$ are then inserted into the
gauge fields $A_0(t,z)$ $A_3(t,z)$ from eqs.\ (\ref{Atz}) to obtain the final solution
\begin{subequations} 
\bea
A_0(t,z) &=& \mu_t-\mu_{5,t}\frac{\sinh(2\beta\arctan z)}{\sinh\beta\pi} \non
&& -(\jmath_t-\epsilon t) \left[\frac{\cosh(2\beta\arctan z)}{\sinh\beta\pi} -\coth\beta\pi\right] \, , \\
A_3(t,z) &=& -tE-\mu_{5,t}\left[\frac{\cosh(2\beta\arctan z)}{\sinh\beta\pi}-\coth\beta\pi\right] \non 
&& -(\jmath_t-\epsilon t) \frac{\sinh(2\beta\arctan z)}{\sinh\beta\pi}  \, .
\eea
\end{subequations}
For the free energy we also need the field strengths (times $k(z)$),
\begin{subequations} 
\bea
k\partial_z A_0 &=& -2\beta\left[\mu_{5,t}\frac{\cosh(2\beta\arctan z)}{\sinh\beta\pi}
+(\jmath_t-\epsilon t)\frac{\sinh(2\beta\arctan z)}{\sinh\beta\pi}\right] \, , \\
k\partial_z A_3 &=& -2\beta\left[\mu_{5,t}\frac{\sinh(2\beta\arctan z)}{\sinh\beta\pi}
+(\jmath_t-\epsilon t)\frac{\cosh(2\beta\arctan z)}{\sinh\beta\pi}\right] \, .
\eea
\end{subequations}
As a check, we can perform a parity transformation on the gauge fields. With $\mu\to+\mu$, $\mu_5\to-\mu_5$, $\jmath_t\to+\jmath_t$, 
$B\to +B$, $E\to -E$, $\epsilon\to +\epsilon$ and $z\to -z$ we find $A_0(t,z)\to A_0(t,z)$ and $A_3(t,z)\to -A_3(t,z)$, i.e., 
the fields have the correct behavior under parity transformations for each $t$ and $z$. 

We can now insert the gauge fields and field strengths into the action (\ref{action}) to obtain the thermodynamic potential 
$\Omega=\frac{T}{V}S_{\rm on-shell}$. The YM and CS contributions are
\begin{subequations}\label{YMCSonshell}
\bea
\Omega_{\rm YM} &=& \kappa M_{\rm KK}^2\frac{4\pi\beta^2}{\sinh^2\beta\pi}\left[(\jmath_t-\epsilon t)^2-\mu_{5,t}^2\right] \, , \\
\Omega_{\rm CS} &=& \frac{8\kappa M_{\rm KK}^2}{3}\left(\beta\coth\beta\pi-\frac{\pi\beta^2}{\sinh^2\beta\pi}\right)
\left[(\jmath_t-\epsilon t)^2-\mu_{5,t}^2\right]\non
&&+\frac{8\kappa M_{\rm KK}^2}{3}\beta\left[\mu_t(\jmath_t-\epsilon t) + tE\mu_{5,t}\right] \, , \label{CSonshell}
\eea
where the real time parameter $t$ is treated as an external parameter,
unrelated to the imaginary time $\tau$, whose integration is assumed
to just give a factor $1/T$.
\end{subequations}
In the YM part we have dropped the terms $\propto B^2, E^2$. This vacuum subtraction
can be understood in terms of holographic renormalization and follows from the renormalization condition that the thermodynamic potential 
be zero for vanishing chemical potentials; for the explicit procedure see ref.\ \cite{Rebhan:2008ur}.

\subsection{Chirally symmetric phase}
\label{app:restored}

Here we solve the equations of motion for the chirally symmetric phase, eqs.\ (\ref{E123sym}). For notational convenience, let us,
in this subsection, denote 
\be
\beta'\equiv \frac{\beta}{\theta^3} \, .
\ee
The time-dependence of the gauge fields is treated analogously to the broken phase. Thus, eqs.\ (\ref{E3sym}) and (\ref{E2sym})
imply 
\be \label{k01}
k_0\partial_z A_0^{L/R} = \mp 2\beta' t e_{L/R}(z) + k_0\partial_z \widetilde{A}_0^{L/R} \, , 
\ee
and 
\be \label{dzA31}
\partial_z A_3^{L/R} = -t\partial_z e_{L/R}(z)+\partial_z \widetilde{A}_3^{L/R} \, ,
\ee
where $\widetilde{A}_0^{L/R}$, $\widetilde{A}_3^{L/R}$ are constant in $t$, and where $e_{L/R}\equiv -\partial_t A_3^{L/R}$. 
Then, eqs.\ (\ref{E1sym}) and (\ref{E2sym}) read 
\begin{subequations} \label{eom5} 
\bea
\partial_z(k_0\partial_z\widetilde{A}_0^{L/R})&=&\pm 2\beta'\partial_z \widetilde{A}_3^{L/R} \, ,\\
\partial_z(k_3\partial_z \widetilde{A}_3^{L/R})&=&\pm 2\beta'\partial_z \widetilde{A}_0^{L/R} 
-t\left[(2\beta')^2\frac{e_{L/R}(z)}{k_0(z)}-\partial_z(k_3\partial_z e_{L/R})\right] 
\label{Er2} \, .
\eea
\end{subequations}
This is analogous to eq.\ (\ref{eom2}), the only difference being the two functions $k_0(z)$ and $k_3(z)$ instead of the single function $k(z)$. 
Again the square bracket in eq.\ (\ref{Er2}) has to vanish. This yields a differential equation for $e_{L/R}(z)$ which is solved as follows.
With $\widetilde{e}_{L/R}=k_3\partial_z e_{L/R}$ one can rewrite this differential equation as 
\be \label{diffeq1}
\partial_z(k_0\partial_z\widetilde{e}_{L/R})=(2\beta')^2\frac{\widetilde{e}_{L/R}}{k_3} \, .
\ee
This equation has the two independent solutions 
\begin{subequations} \label{pq}
\bea
p(z) &=& {}_2F_1\left[-\frac{\sqrt{1-16\beta'^2}+1}{4},\frac{\sqrt{1-16\beta'^2}-1}{4},\frac{1}{2},
\frac{1}{1+z^2}\right] \, , \\
q(z) &=& \frac{1}{\sqrt{1+z^2}}\, {}_2F_1\left[-\frac{\sqrt{1-16\beta'^2}-1}{4},\frac{\sqrt{1-16\beta'^2}+1}{4},\frac{3}{2},
\frac{1}{1+z^2}\right] \, .
\eea
\end{subequations}
Consequently, $\widetilde{e}_{L/R}(z)=P_{L/R}\,p(z)+Q_{L/R}\,q(z)$, with constants $P_{L/R}$, $Q_{L/R}$, and thus
\be \label{ecc}
e_{L/R}(z) = \frac{1}{(2\beta')^2}\left(P_{L/R}\, k_0\partial_zp+Q_{L/R}\,k_0\partial_zq\right) \, .
\ee
In the following we need the behavior of the functions $p(z)$, $q(z)$, $k_0\partial_zp$, $k_0\partial_zq$ 
at $z=\infty$ and $z=0$. At $z=\infty$ we have 
\be \label{tauinfty}
p(\infty) = -k_0\partial_zq(\infty) = 1 \, , \qquad q(\infty) = k_0\partial_zp(\infty) = 0 \, .
\ee
At $z=0$ one finds 
\begin{subequations} \label{tauzero1}
\bea
p_0&\equiv&
p(0) = \frac{\sqrt{\pi}}{\Gamma\left[\left(3-\sqrt{1-16\beta'^2}\right)/4\right]\Gamma\left[\left(3+\sqrt{1-16\beta'^2}\right)/4
\right]} \, , \\
q_0&\equiv& q(0) = \frac{\sqrt{\pi}}{2\Gamma\left[\left(5-\sqrt{1-16\beta'^2}\right)/4\right]
\Gamma\left[\left(5+\sqrt{1-16\beta'^2}\right)/4
\right]} \, , 
\eea
\end{subequations}
and
\be \label{tauzero2}
k_0\partial_z p (z\to 0) = (2\beta')^2 p_0 \,\ln z \, , \qquad k_0\partial_z q (z\to 0) = (2\beta')^2 q_0 \, \ln z\, .
\ee
The boundary conditions $e_{L/R}(z=\infty) = E\mp\epsilon$
yield $Q_{L/R}=-(2\beta')^2(E\mp\epsilon)$. Inserting this constant into eq.\ (\ref{ecc}), the result
into eqs.\ (\ref{k01}), (\ref{dzA31}), and integrating the resulting equations over $z$ yields the gauge fields
\begin{subequations} \label{Atzsym}
\bea
A_0^{L/R}(t,z)&=&\mp2\beta' t\left[\frac{P_{L/R}}{(2\beta')^2}\,p(z)-(E\mp\epsilon)\,q(z)\right] + g_0^{L/R}(t) + \widetilde{A}_0^{L/R}(z) 
\, , \\
A_3^{L/R}(t,z)&=&-t\left[\frac{P_{L/R}}{(2\beta')^2}\,k_0\partial_z p(z)-(E\mp\epsilon)\,k_0\partial_z q\right] 
+ \widetilde{A}_3^{L/R}(z) 
\, .
\eea
\end{subequations}
Here, $g_0^{L/R}(t)$ are time-dependent integration constants from the $z$ integration. 
We proceed by solving eqs.\ (\ref{eom5}) for $\widetilde{A}_0^{L/R}$, $\widetilde{A}_3^{L/R}$. 
Recalling that $p(z)$, $q(z)$ fulfill the differential equation (\ref{diffeq1}) one easily checks that
the functions
\begin{subequations} \label{gauge1}
\bea
\widetilde{A}_0^{L/R}(z) &=& a_0^{L/R} \pm 2\beta' \left[ C_{L/R}\, p(z) + D_{L/R}\, q(z)\right] \, , \\
\widetilde{A}_3^{L/R}(z) &=& a_3^{L/R} + C_{L/R}\, k_0\partial_z p + D_{L/R}\,k_0 \partial_z q \, , 
\eea
\end{subequations}
with integration constants $a_0^{L/R}$, $a_3^{L/R}$, $C_{L/R}$ and $D_{L/R}$, are solutions of eqs.\ (\ref{eom5}). 
One now inserts these functions into eqs.\ (\ref{Atzsym}) and determines 
the integration constants as follows. First we recall that all constants except for $g_0^{L/R}(t)$ must not depend on $t$.
This will be used repeatedly in the following. 
Then we require the boundary condition $\mbox{$A_3^{L/R}(t,z=\infty)$}=-t(E\mp\epsilon)$ which implies $D_{L/R} = a_3^{L/R}$. Next, we
require regularity of $A_3^{L/R}(t,z)$ at $z=0$. With eq.\ (\ref{tauzero2}) we find that $A_3^{L/R}(t,z\to 0)$ diverges logarithmically.
Requiring the factor in front of the $\ln z$ term to vanish yields the conditions
\be
C_{L/R} = -\frac{q_0}{p_0} D_{L/R} \, , \qquad P_{L/R} = (2\beta')^2\frac{q_0}{p_0}(E\mp\epsilon) \, .
\ee
For the temporal component we need to require $A_0^{L/R}(t,z=0)=0$ \cite{Horigome:2006xu} which yields $a_0^{L/R} = g_0^{L/R}(t) = 0$. 
With these results the boundary value of $A_0^{L/R}(t,z)$ becomes
\be
A_0^{L/R}(t,z=\infty) = \mp2\beta'\frac{q_0}{p_0}[D_{L/R}+t(E\mp\epsilon)] \, .
\ee
This result shows that, as in the broken phase, the boundary values of 
axial and vector parts of $A_0$ necessarily become time-dependent. In other
words, in the presence of an electric field one cannot fix these boundary values to be time-independent chemical potentials. 
At $t=0$ we require $A_0^{L/R}(t=0,z=\infty)=\mu\mp\mu_5$. With these initial values we find
\be
D_{L/R} = \mp \frac{p_0}{q_0}\frac{\mu\mp\mu_5}{2\beta'} \, , 
\ee
and the time-dependent chemical potentials become
\be
A_0^{L/R}(t,z=\infty) = \mu_t\mp\mu_{5,t} \, , 
\ee
with  
\be \label{mut}
\mu_t \equiv \mu+2\beta' t\epsilon\frac{q_0}{p_0} \, , \qquad
\mu_{5,t} \equiv \mu_5+2\beta' tE\frac{q_0}{p_0} \, .
\ee
Collecting all the integration constants, we obtain from eqs.\ (\ref{Atzsym}) and (\ref{gauge1}) the final solution for the gauge 
fields,
\begin{subequations} 
\bea
A_0^{L/R}(t,z) &=& (\mu_t\mp\mu_{5,t})\left[p(z)-\frac{p_0}{q_0}\,q(z)\right] \, , \\
A_3^{L/R}(t,z) &=& -t(E\mp \epsilon)\pm\frac{\mu_t\mp\mu_{5,t}}{2\beta'}
\left[k_0\partial_zp-\frac{p_0}{q_0}(1+k_0\partial_zq)\right]
\, .
\eea
\end{subequations}
Again we can check the behavior of the gauge fields 
under parity transformations. In contrast to the broken phase, we have separate right- and left-handed fields which transform as 
$A_0^{L/R}(t,z)\to A_0^{R/L}(t,z)$, and $A_3^{L/R}(t,z)\to - A_3^{R/L}(t,z)$, as it should be.  
The field strengths become
\begin{subequations} 
\bea
k_0\partial_z A_0^{L/R} &=&(\mu_t\mp\mu_{5,t})
\left(k_0\partial_zp-\frac{p_0}{q_0}k_0\partial_zq\right) 
\, , \\
k_3\partial_z A_3^{L/R} &=& \pm 2\beta' (\mu_t\mp\mu_{5,t})\left[p(z)-\frac{p_0}{q_0}q(z)\right] \, .
\eea
\end{subequations}
Inserting these results into the action, given by eqs.\ (\ref{SRL}) and (\ref{actionsym}), yields the 
YM and CS contributions to the free energy, 
\begin{subequations} \label{YMCSOmega2}
\bea
\Omega_{\rm YM} &=& -2\kappa\theta^3 M_{\rm KK}^2(\mu_t^2+\mu_{5,t}^2)[I_0-(2\beta')^2I_3] \, , \\
\Omega_{\rm CS} &=& \frac{4\kappa M_{\rm KK}^2\theta^3}{3}
\Big\{(\mu_t^2+\mu_{5,t}^2)\Big[I_0-(2\beta')^2I_3-\frac{p_0}{q_0}\Big] +2\beta'\,t\left(\mu_t\epsilon+ \mu_{5,t} E\right)
\Big\} \, ,
\eea
\end{subequations}
where we abbreviated the integrals 
\bea \label{defI}
I_0&\equiv & \int_0^\infty \frac{dz}{k_0}\left(k_0\partial_zp-\frac{p_0}{q_0} k_0\partial_zq\right)^2 \, , \\
I_3&\equiv & \int_0^\infty \frac{dz}{k_3}\left[p(z)-\frac{p_0}{q_0} q(z)\right]^2 \, .
\eea
In the limit $\beta\gg 1$, the combination $I_0-(2\beta')^2I_3\to 0$, so that for very strong magnetic fields $\Omega_{\rm CS}\gg \Omega_{\rm YM}$, as is also
the case in the chirally broken phase, see eqs.~(\ref{YMCSonshell}).

\bibliographystyle{JHEP}
\bibliography{refs1}

\providecommand{\href}[2]{#2}\begingroup\raggedright\begin{thebibliography}{10}

\bibitem{Kharzeev:2007jp}
D.~E. Kharzeev, L.~D. McLerran, and H.~J. Warringa, {\it {The effects of
  topological charge change in heavy ion collisions: 'Event by event P and CP
  violation'}},  {\em Nucl. Phys.} {\bf A803} (2008) 227--253,
  [\href{http://arxiv.org/abs/0711.0950}{{\tt arXiv:0711.0950}}].

\bibitem{Fukushima:2008xe}
K.~Fukushima, D.~E. Kharzeev, and H.~J. Warringa, {\it {The Chiral Magnetic
  Effect}},  {\em Phys. Rev.} {\bf D78} (2008) 074033,
  [\href{http://arxiv.org/abs/0808.3382}{{\tt arXiv:0808.3382}}].

\bibitem{Kharzeev:2009pj}
D.~E. Kharzeev and H.~J. Warringa, {\it {Chiral Magnetic conductivity}},  {\em
  Phys. Rev.} {\bf D80} (2009) 034028,
  [\href{http://arxiv.org/abs/0907.5007}{{\tt arXiv:0907.5007}}].

\bibitem{Voloshin:2008jx}
{\bf STAR} Collaboration, S.~A. Voloshin, {\it {Probe for the strong parity
  violation effects at RHIC with three particle correlations}},
  \href{http://arxiv.org/abs/0806.0029}{{\tt arXiv:0806.0029}}.

\bibitem{Abelev:2009uh}
{\bf STAR} Collaboration, B.~I. Abelev {\em et~al.}, {\it {Azimuthal
  Charged-Particle Correlations and Possible Local Strong Parity Violation}},
  \href{http://arxiv.org/abs/0909.1739}{{\tt arXiv:0909.1739}}.

\bibitem{Wang:2009kd}
F.~Wang, {\it {Effects of Cluster Particle Correlations on Local Parity
  Violation Observables}},  \href{http://arxiv.org/abs/0911.1482}{{\tt
  arXiv:0911.1482}}.

\bibitem{Yee:2009vw}
H.-U. Yee, {\it {Holographic Chiral Magnetic Conductivity}},
  \href{http://arxiv.org/abs/0908.4189}{{\tt arXiv:0908.4189}}.

\bibitem{Son:2004tq}
D.~T. Son and A.~R. Zhitnitsky, {\it {Quantum anomalies in dense matter}},
  {\em Phys. Rev.} {\bf D70} (2004) 074018,
  [\href{http://arxiv.org/abs/hep-ph/0405216}{{\tt hep-ph/0405216}}].

\bibitem{Metlitski:2005pr}
M.~A. Metlitski and A.~R. Zhitnitsky, {\it {Anomalous axion interactions and
  topological currents in dense matter}},  {\em Phys. Rev.} {\bf D72} (2005)
  045011, [\href{http://arxiv.org/abs/hep-ph/0505072}{{\tt hep-ph/0505072}}].

\bibitem{Gorbar:2009bm}
E.~V. Gorbar, V.~A. Miransky, and I.~A. Shovkovy, {\it {Chiral asymmetry of the
  Fermi surface in dense relativistic matter in a magnetic field}},  {\em Phys.
  Rev.} {\bf C80} (2009) 032801, [\href{http://arxiv.org/abs/0904.2164}{{\tt
  arXiv:0904.2164}}].

\bibitem{Charbonneau:2009ax}
J.~Charbonneau and A.~Zhitnitsky, {\it {Topological Currents in Neutron Stars:
  Kicks, Precession, Toroidal Fields, and Magnetic Helicity}},
  \href{http://arxiv.org/abs/0903.4450}{{\tt arXiv:0903.4450}}.

\bibitem{Maldacena:1997re}
J.~M. Maldacena, {\it {The large N limit of superconformal field theories and
  supergravity}},  {\em Adv. Theor. Math. Phys.} {\bf 2} (1998) 231--252,
  [\href{http://arxiv.org/abs/hep-th/9711200}{{\tt hep-th/9711200}}].

\bibitem{Gubser:1998bc}
S.~S. Gubser, I.~R. Klebanov, and A.~M. Polyakov, {\it {Gauge theory
  correlators from non-critical string theory}},  {\em Phys. Lett.} {\bf B428}
  (1998) 105--114, [\href{http://arxiv.org/abs/hep-th/9802109}{{\tt
  hep-th/9802109}}].

\bibitem{Witten:1998qj}
E.~Witten, {\it {Anti-de Sitter space and holography}},  {\em Adv. Theor. Math.
  Phys.} {\bf 2} (1998) 253--291,
  [\href{http://arxiv.org/abs/hep-th/9802150}{{\tt hep-th/9802150}}].

\bibitem{Alford:1998mk}
M.~G. Alford, K.~Rajagopal, and F.~Wilczek, {\it {Color-flavor locking and
  chiral symmetry breaking in high density {QCD}}},  {\em Nucl. Phys.} {\bf
  B537} (1999) 443--458, [\href{http://arxiv.org/abs/hep-ph/9804403}{{\tt
  hep-ph/9804403}}].

\bibitem{Sakai:2004cn}
T.~Sakai and S.~Sugimoto, {\it {Low energy hadron physics in holographic QCD}},
   {\em Prog. Theor. Phys.} {\bf 113} (2005) 843--882,
  [\href{http://arxiv.org/abs/hep-th/0412141}{{\tt hep-th/0412141}}].

\bibitem{Sakai:2005yt}
T.~Sakai and S.~Sugimoto, {\it {More on a holographic dual of QCD}},  {\em
  Prog. Theor. Phys.} {\bf 114} (2005) 1083--1118,
  [\href{http://arxiv.org/abs/hep-th/0507073}{{\tt hep-th/0507073}}].

\bibitem{Bergman:2008qv}
O.~Bergman, G.~Lifschytz, and M.~Lippert, {\it {Magnetic properties of dense
  holographic QCD}},  {\em Phys. Rev.} {\bf D79} (2009) 105024,
  [\href{http://arxiv.org/abs/0806.0366}{{\tt arXiv:0806.0366}}].

\bibitem{Thompson:2008qw}
E.~G. Thompson and D.~T. Son, {\it {Magnetized baryonic matter in holographic
  QCD}},  {\em Phys. Rev.} {\bf D78} (2008) 066007,
  [\href{http://arxiv.org/abs/0806.0367}{{\tt arXiv:0806.0367}}].

\bibitem{Rebhan:2008ur}
A.~Rebhan, A.~Schmitt, and S.~A. Stricker, {\it {Meson supercurrents and the
  Meissner effect in the Sakai-Sugimoto model}},  {\em JHEP} {\bf 05} (2009)
  084, [\href{http://arxiv.org/abs/0811.3533}{{\tt arXiv:0811.3533}}].

\bibitem{Lifschytz:2009si}
G.~Lifschytz and M.~Lippert, {\it {Anomalous conductivity in holographic QCD}},
   {\em Phys. Rev.} {\bf D80} (2009) 066005,
  [\href{http://arxiv.org/abs/0904.4772}{{\tt arXiv:0904.4772}}].

\bibitem{Buividovich:2009wi}
P.~V. Buividovich, M.~N. Chernodub, E.~V. Luschevskaya, and M.~I. Polikarpov,
  {\it {Numerical evidence of chiral magnetic effect in lattice gauge theory}},
   \href{http://arxiv.org/abs/0907.0494}{{\tt arXiv:0907.0494}}.

\bibitem{Hata:2008xc}
H.~Hata, M.~Murata, and S.~Yamato, {\it {Chiral currents and static properties
  of nucleons in holographic QCD}},  {\em Phys. Rev.} {\bf D78} (2008) 086006,
  [\href{http://arxiv.org/abs/0803.0180}{{\tt arXiv:0803.0180}}].

\bibitem{McLerran:2007qj}
L.~McLerran and R.~D. Pisarski, {\it {Phases of Cold, Dense Quarks at Large
  $N_c$}},  {\em Nucl. Phys.} {\bf A796} (2007) 83--100,
  [\href{http://arxiv.org/abs/0706.2191}{{\tt arXiv:0706.2191}}].

\bibitem{Hata:2007mb}
H.~Hata, T.~Sakai, S.~Sugimoto, and S.~Yamato, {\it {Baryons from instantons in
  holographic QCD}},  \href{http://arxiv.org/abs/hep-th/0701280}{{\tt
  hep-th/0701280}}.

\bibitem{Hashimoto:2008zw}
K.~Hashimoto, T.~Sakai, and S.~Sugimoto, {\it {Holographic Baryons : Static
  Properties and Form Factors from Gauge/String Duality}},  {\em Prog. Theor.
  Phys.} {\bf 120} (2008) 1093--1137,
  [\href{http://arxiv.org/abs/0806.3122}{{\tt arXiv:0806.3122}}].

\bibitem{Kim:2008pw}
K.-Y. Kim and I.~Zahed, {\it {Electromagnetic Baryon Form Factors from
  Holographic QCD}},  {\em JHEP} {\bf 09} (2008) 007,
  [\href{http://arxiv.org/abs/0807.0033}{{\tt arXiv:0807.0033}}].

\bibitem{Kim:2008iy}
K.-Y. Kim and I.~Zahed, {\it {Nucleon-Nucleon Potential from Holography}},
  {\em JHEP} {\bf 03} (2009) 131, [\href{http://arxiv.org/abs/0901.0012}{{\tt
  arXiv:0901.0012}}].

\bibitem{Bardeen:1969md}
W.~A. Bardeen, {\it {Anomalous Ward identities in spinor field theories}},
  {\em Phys. Rev.} {\bf 184} (1969) 1848--1857.

\bibitem{Hill:2006ei}
C.~T. Hill, {\it {Anomalies, Chern-Simons terms and chiral delocalization in
  extra dimensions}},  {\em Phys. Rev.} {\bf D73} (2006) 085001,
  [\href{http://arxiv.org/abs/hep-th/0601154}{{\tt hep-th/0601154}}].

\bibitem{Bell:1969ts}
J.~S. Bell and R.~Jackiw, {\it {A PCAC puzzle: $\pi_0 \to \gamma \gamma$ in the
  sigma model}},  {\em Nuovo Cim.} {\bf A60} (1969) 47--61.

\bibitem{Adler:1969gk}
S.~L. Adler, {\it {Axial vector vertex in spinor electrodynamics}},  {\em Phys.
  Rev.} {\bf 177} (1969) 2426--2438.

\bibitem{Kaymakcalan:1983qq}
O.~Kaymakcalan, S.~Rajeev, and J.~Schechter, {\it {Nonabelian Anomaly and
  Vector Meson Decays}},  {\em Phys. Rev.} {\bf D30} (1984) 594.

\bibitem{Erdmenger:2008rm}
J.~Erdmenger, M.~Haack, M.~Kaminski, and A.~Yarom, {\it {Fluid dynamics of
  R-charged black holes}},  {\em JHEP} {\bf 01} (2009) 055,
  [\href{http://arxiv.org/abs/0809.2488}{{\tt arXiv:0809.2488}}].

\bibitem{Banerjee:2008th}
N.~Banerjee, J.~Bhattacharya, S.~Bhattacharyya, S.~Dutta, R.~Loganayagam, and
  P.~Surowka, {\it {Hydrodynamics from charged black branes}},
  \href{http://arxiv.org/abs/0809.2596}{{\tt arXiv:0809.2596}}.

\bibitem{Torabian:2009qk}
M.~Torabian and H.-U. Yee, {\it {Holographic nonlinear hydrodynamics from
  AdS/CFT with multiple/non-Abelian symmetries}},  {\em JHEP} {\bf 08} (2009)
  020, [\href{http://arxiv.org/abs/0903.4894}{{\tt arXiv:0903.4894}}].

\bibitem{Son:2009tf}
D.~T. Son and P.~Surowka, {\it {Hydrodynamics with Triangle Anomalies}},
  \href{http://arxiv.org/abs/0906.5044}{{\tt arXiv:0906.5044}}.

\bibitem{Bergman:2008sg}
O.~Bergman, G.~Lifschytz, and M.~Lippert, {\it {Response of Holographic QCD to
  Electric and Magnetic Fields}},  {\em JHEP} {\bf 05} (2008) 007,
  [\href{http://arxiv.org/abs/0802.3720}{{\tt arXiv:0802.3720}}].

\bibitem{Johnson:2008vna}
C.~V. Johnson and A.~Kundu, {\it {External Fields and Chiral Symmetry Breaking
  in the Sakai-Sugimoto Model}},  {\em JHEP} {\bf 12} (2008) 053,
  [\href{http://arxiv.org/abs/0803.0038}{{\tt arXiv:0803.0038}}].

\bibitem{Kim:2007vd}
K.-Y. Kim, S.-J. Sin, and I.~Zahed, {\it {Dense Holographic QCD in the
  Wigner-Seitz Approximation}},  {\em JHEP} {\bf 09} (2008) 001,
  [\href{http://arxiv.org/abs/0712.1582}{{\tt arXiv:0712.1582}}].

\bibitem{Ambjorn:1983hp}
J.~Ambjorn, J.~Greensite, and C.~Peterson, {\it {The axial anomaly and the
  lattice Dirac sea}},  {\em Nucl. Phys.} {\bf B221} (1983) 381.

\bibitem{Lifschytz:2009sz}
G.~Lifschytz and M.~Lippert, {\it {Holographic Magnetic Phase Transition}},
  {\em Phys. Rev.} {\bf D80} (2009) 066007,
  [\href{http://arxiv.org/abs/0906.3892}{{\tt arXiv:0906.3892}}].

\bibitem{Elitzur:1989nr}
S.~Elitzur, G.~W. Moore, A.~Schwimmer, and N.~Seiberg, {\it {Remarks on the
  Canonical Quantization of the Chern-Simons-Witten Theory}},  {\em Nucl.
  Phys.} {\bf B326} (1989) 108.

\bibitem{Newman:2005as}
G.~M. Newman and D.~T. Son, {\it {Response of strongly-interacting matter to
  magnetic field: Some exact results}},  {\em Phys. Rev.} {\bf D73} (2006)
  045006, [\href{http://arxiv.org/abs/hep-ph/0510049}{{\tt hep-ph/0510049}}].

\bibitem{Nielsen:1983rb}
H.~B. Nielsen and M.~Ninomiya, {\it {Adler-Bell-Jackiw anomaly and Weyl
  fermions in crystal}},  {\em Phys. Lett.} {\bf B130} (1983) 389.

\bibitem{Basu:2009qz}
P.~Basu, J.~He, A.~Mukherjee, and H.-H. Shieh, {\it {Holographic Non-Fermi
  Liquid in a Background Magnetic Field}},
  \href{http://arxiv.org/abs/0908.1436}{{\tt arXiv:0908.1436}}.

\bibitem{Denef:2009yy}
F.~Denef, S.~A. Hartnoll, and S.~Sachdev, {\it {Quantum oscillations and black
  hole ringing}},  \href{http://arxiv.org/abs/0908.1788}{{\tt
  arXiv:0908.1788}}.

\bibitem{priv}
H.-U. Yee, {\it {private communication}},  2009.

\bibitem{Horigome:2006xu}
N.~Horigome and Y.~Tanii, {\it {Holographic chiral phase transition with
  chemical potential}},  {\em JHEP} {\bf 01} (2007) 072,
  [\href{http://arxiv.org/abs/hep-th/0608198}{{\tt hep-th/0608198}}].

\end{thebibliography}\endgroup

\end{document}